\documentclass[aps, pre, showpacs, notitlepage]{revtex4-1}
\usepackage{graphicx}
\usepackage{float}
\usepackage{amssymb,amsmath}
\usepackage{color}
\usepackage{bm}
\usepackage{subfigure}
\usepackage{ulem} 
\graphicspath{{./images/}}

\begin{document}

\normalem 

\title{Noise in Coevolving Networks}
\author{Marina Diakonova}
\author{V\'ictor M. Egu\'iluz}
\author{Maxi San Miguel}
\affiliation{Instituto de F\'isica Interdisciplinar y Sistemas Complejos IFISC (CSIC-UIB), E07122 Palma de Mallorca, Spain}

\begin{abstract}
Coupling dynamics of the states of the nodes of a network to the dynamics of the network topology leads to generic absorbing and fragmentation transitions. The coevolving voter model is a typical system that exhibits such transitions at some critical rewiring. We study the robustness of these transitions under two distinct ways of introducing noise. Noise affecting all the nodes destroys the absorbing-fragmentation transition, giving rise in finite-size systems to two regimes: bimodal magnetisation and dynamic fragmentation. Noise Targeting a fraction of nodes preserves the transitions but introduces shattered fragmentation with its characteristic fraction of isolated nodes and one or two giant components. Both the lack of absorbing state for homogenous noise and the shift in the absorbing transition to higher rewiring for targeted noise are supported by analytical approximations.
\end{abstract}

\maketitle
\section{Introduction} 
\label{sec:introduction}
Coevolving, or adaptive, networks are a prominent framework in complexity science that models a range of phenomena where the state of node is influenced by, and simultaneously shapes, the neighbourhood structure \cite{Gross2008,Castellano2009,Gross2009}. This coupling of the dynamics \emph{on} the network to the dynamics \emph{of} the network leads to new, qualitatively different long term behaviour of the system \cite{Zimmermann2001,Zimmermann2004, Vazquez2007, Vazquez2008,Gross2008,Castellano2009}. A generic feature of coevolving networks is the fragmentation transition occurring at some critical value of the plasticity parameter, which relates the preference for topological evolution over changes of state of the nodes. Networks with smaller plasticities tend to remain connected, while larger plasticities lead to the coexistence of several disconnected communities with differing states \cite{Nardini2008, Bohme2012, Singh2012, Shi2013, Nicosia2013, Malik2013, Su2014, Benczik2009, Gil2006, Holme2006, Kimura2008, Lazer2010, Herrera2011}.

Noise is expected to be intrinsic to real systems, however, it has been largely unexplored in coevolution models. Attempts to take into account noise effects include studying epidemic spreading \cite{Shaw2009} with noise terms in the stochastic mean-field model, looking at robustness of fragmented clusters in the diffusion of cultural traits \cite{Centola2007}, as well as considering how noise promotes community structure \cite{JuanCarlos2014}. Adding a small amount of noise to adaptive networks has also been shown to stabilize them enough to be used as testing ground for analytical methodologies \cite{Ji2013}, \textcolor{black}{while a model of epidemic spreading with noise on the links was successfully used to develop an alternative approximative modelling approach \cite{Galla2012}.} In this paper we address the question of the effect of noise on the phase transitions found on coevolution models, namely the robustness or possible modification of fragmentation transitions.  

In the context where agents' states and connections are interrelated, noise models fluctuations that can be considered as either internal or external, but in any case not related to the general state of the system. We restrict here to noise acting as random changes in the states of nodes. An `internal' interpretation of this type of noise, in the context of collective social phenomena, is that of `free will', making nodes that are subject to noise more individual and less controlled by their social contexts. Alternatively, such nodes can represent individuals that are all-too-easily responsive to outside influence like the media, and are constantly swayed by it. These susceptible individuals can be thought of as the opposite of zealots, which are modelled as nodes that never change their state. The issue of zealots raises questions about how their presence and numbers affects the state of the system \cite{Mobilia2007}. Our investigations can thus be considered as the complementary problem. We ask how the presence of these suggestible agents alters the transitions observed in standard coevolving networks.

Our starting point is the archetype of such models, the Coevolving Voter Model (the CVM, \cite{Vazquez2008}). It traces the qualitatively different behaviours through the absorbing and fragmentation transitions that occur with the growth of the relative plasticity of the network. In the absence of plasticity, the binary-state voter model changes state of each node to that of its random neighbour. In random networks this system displays a steady level of activity, until finite-size fluctuations ensure the network freezes in one of the two equivalent absorbing states, with all nodes in the same state. Here plasticity is associated with the rewiring probability: increasing the plasticity allows the node to, in place of adopting the state of the neighbour, sever the `active' link joining two discordant nodes and reconnect to a new node that shares its state. There exists a critical value of the plasticity parameter for which an absorbing transition occurs in the thermodynamic limit. This transitions manifest itself in finite systems in a fragmentation beyond which the system ends up in two disjoint network components. The nodes in each of the network components have reached the same state and the two components are in opposite states. This paper investigates the effect of noise on the critical value and the nature of this absorbing and fragmentation transition.

To model a situation where noise may affect some agents less that others, we separate the concept of noise intensity $\epsilon$ from that of the fraction $q$ of the population that is susceptible to it. The framework is defined by the two limiting cases: the first is what we call `homogenous' noise of variable intensity $\epsilon$ that affects all nodes ($q = 1$), and the second a `targeted' noise that affects $q$ nodes at full intensity ($\epsilon = 1$). In this work we investigate these two types of noise, and compare their effects on the absorbing and fragmentation transition.

The work is structured as follows. In section ~\ref{sec:the_model} we introduce the dynamical rules of the model. Homogenous noise is treated in section \ref{sec:homogenous_noise} and the effect of targeting in section \ref{sec:targeted_noise}. We summarize our findings in section \ref{sec:conclusions_extensions}.


\section{The Model} 
\label{sec:the_model}

\begin{figure*}[]
\centering
\hspace*{-0.1 in}
\includegraphics[width = 0.85\textwidth, keepaspectratio = true]{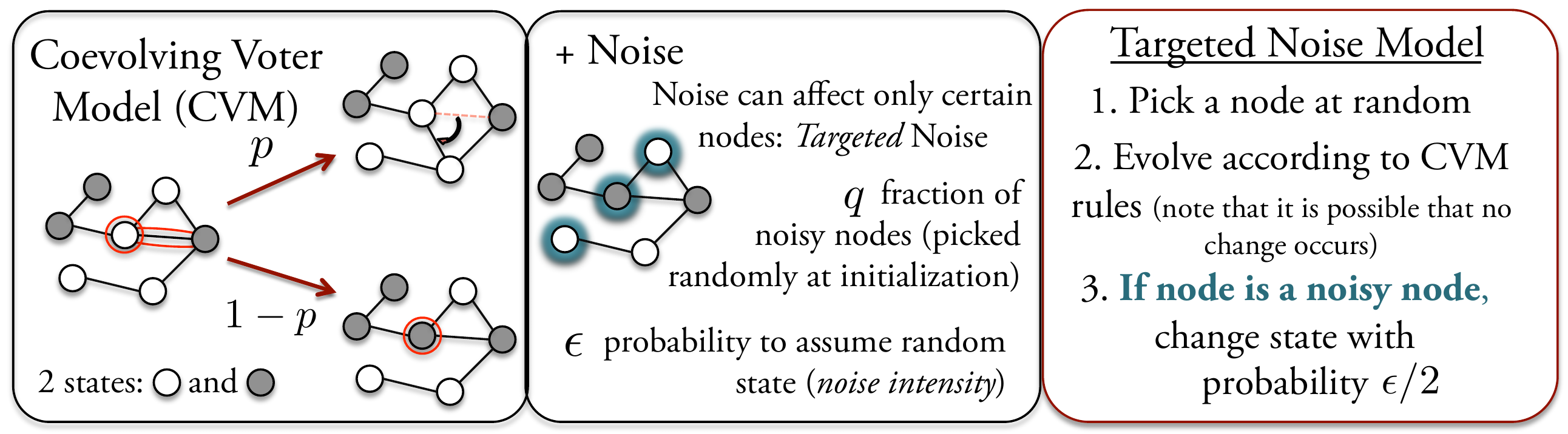}
\caption[]{Schematic illustration of the Targeted Noise Model.}
\label{schema}
\end{figure*}

Our model considers the CVM \cite{Vazquez2008} in a setting that makes some agents susceptible to noise. A schematic illustration is shown in Fig.~\ref{schema}. We start with a network of $N$ agents (nodes) each in one of the two possible states (say, $+1$ and $-1$) that are assigned \textcolor{black}{with equal probability of $1/2$, and independently from each other}. A fraction $q$ of the nodes are then labelled as `noisy', that is, subject to noise. An update proceeds as follows:
\begin{enumerate}
	\item A node $i$ is picked at random. If it has neighbours, then a neighbour $j$ is chosen, also at random. Their states are compared, and if different, then with probability $1-p$ node $i$ changes its state, becoming the same as $j$. Alternatively, with probability $p$, $i$ chooses a random node $k$ from the set of all those that are not connected to it, and are in the same state as $i$. If such a node exists, $i$ severs the link with $j$ and draws a link to $k$ in a process of rewiring.
	\item If node $i$ is `noisy', it assumes a random state with probability $\epsilon$, i.e., changes its state with probability $\epsilon/2$.
\end{enumerate}
There are $N$ such updates in a single timestep. The first step of each update can be recognised as the evolution rule of the CVM, and the second accounts for noise. The model thus has three parameters: rewiring probability $p$, level of targeting expressed through fraction $q$ of `noisy' nodes, and noise intensity $\epsilon$ (note that a `noisy' node will change its state with probability $\epsilon/2$, meaning that $\epsilon = 1$ corresponds to a complete lack of a preferred state). We take as initial condition random regular networks with $N$ nodes and average degree $\mu = 4$, and study system dynamics and configurations in the topologically absorbing state, i.e. one where the network configuration remains fixed for all times. $N$ and $\mu$ are chosen so that the network is initially connected. This choice of network parameters is representative of random regular networks with sparse connectivity, and we expect the phenomena observed in this work to hold for a broad range of $\mu$.

In the CVM, the dynamics is quantified using the density of active links $\rho$, that is, links that join nodes in differing states. In the standard noise-free CVM where the node states depend only on the network and the initial condition, $\rho = 0$ thus corresponds to an absorbing configuration. The topology of these absorbing configurations is characterised by the relative size $S_1$ of the largest network component. For low rewiring probability, $p < p_c$, in the thermodynamic limit the network is active with $\rho > 0$, whereas finite-size systems freeze with the network in consensus, in one giant component with, asymptotically, $S_1 = 1$. The critical rewiring $p_c$ defines the absorbing and fragmentation transitions, beyond which, for $p > p_c$, both the finite and infinite systems are frozen ($\rho = 0$) in two ($S_1 = 0.5$) giant components of opposing states. Figure~\ref{HomNoise:0} shows how the average over realizations of $\rho$, for both $p < p_c$ and $p > p_c$ ($p_c \approx 0.38$), can approximate the asymptotic behaviour and trace the absorbing transition. In this work we will investigate how noise affects the existence and the features of the transition for a critical $p_c$.

To proceed we note a difference in terminology. In our model the noise (step 2) acts irrespective of the outcome of the update (i.e., in step 1). Therefore as long as $q > 0$ and $\epsilon > 0$, \textcolor{black}{some nodes will at some point `spontaneously' change their states; therefore there will no longer be any absorbing configurations of the node states. Depending on $q$, however, the possibility exists of network configurations remaining constant even though state of the nodes might change, and we call those configurations topologically absorbing.}

\section{Homogenous Noise} 
\label{sec:homogenous_noise}

\begin{figure*}
\hspace*{-0.1 in}
\subfigure[System activity]{
\includegraphics[width = 0.35\textwidth]{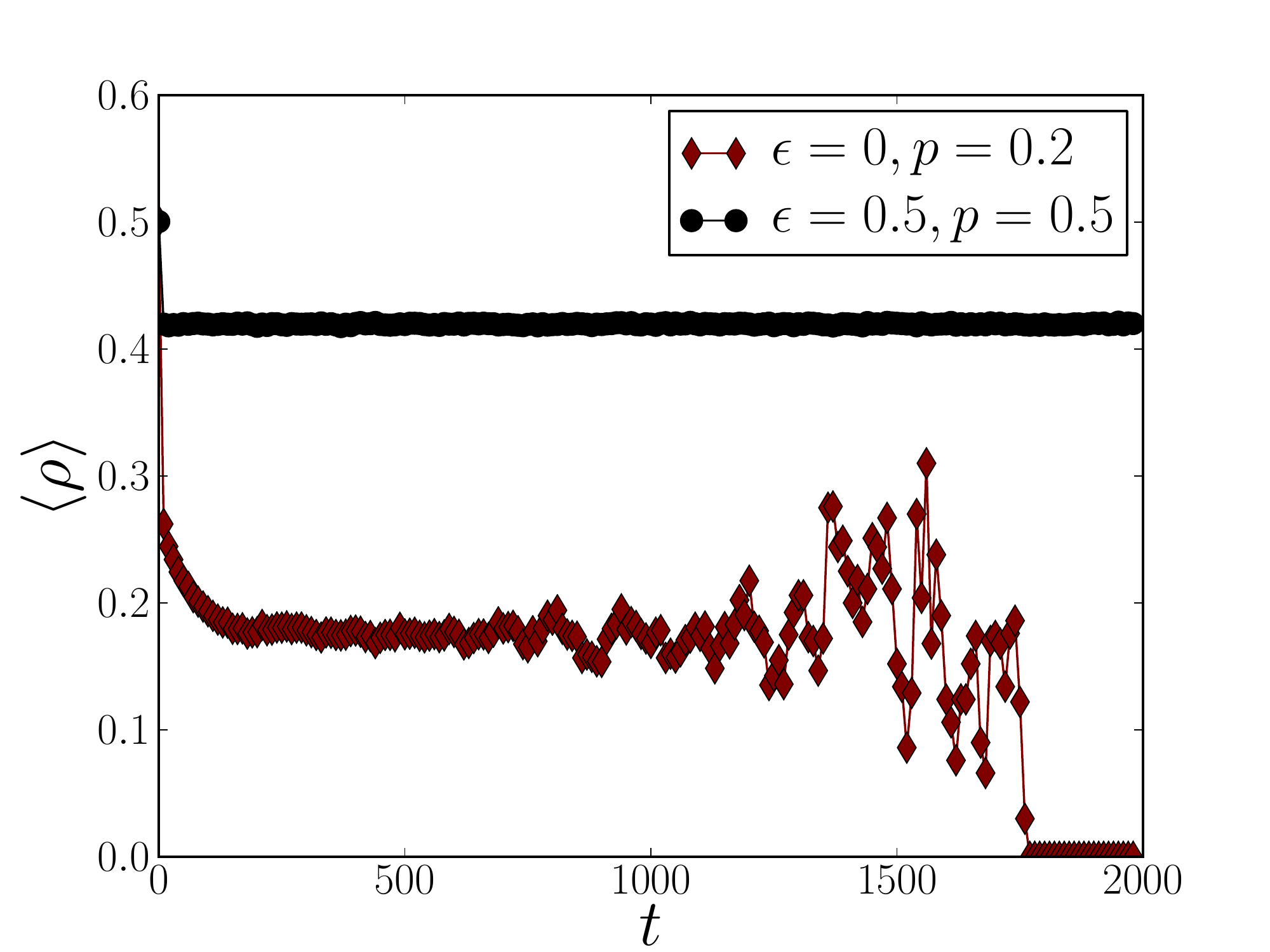}
\label{HomNoise:0}
}
\hspace*{-0.4 in}
\subfigure[Activity for $\epsilon = 0.03$]{
\includegraphics[width = 0.33\textwidth, keepaspectratio = true]{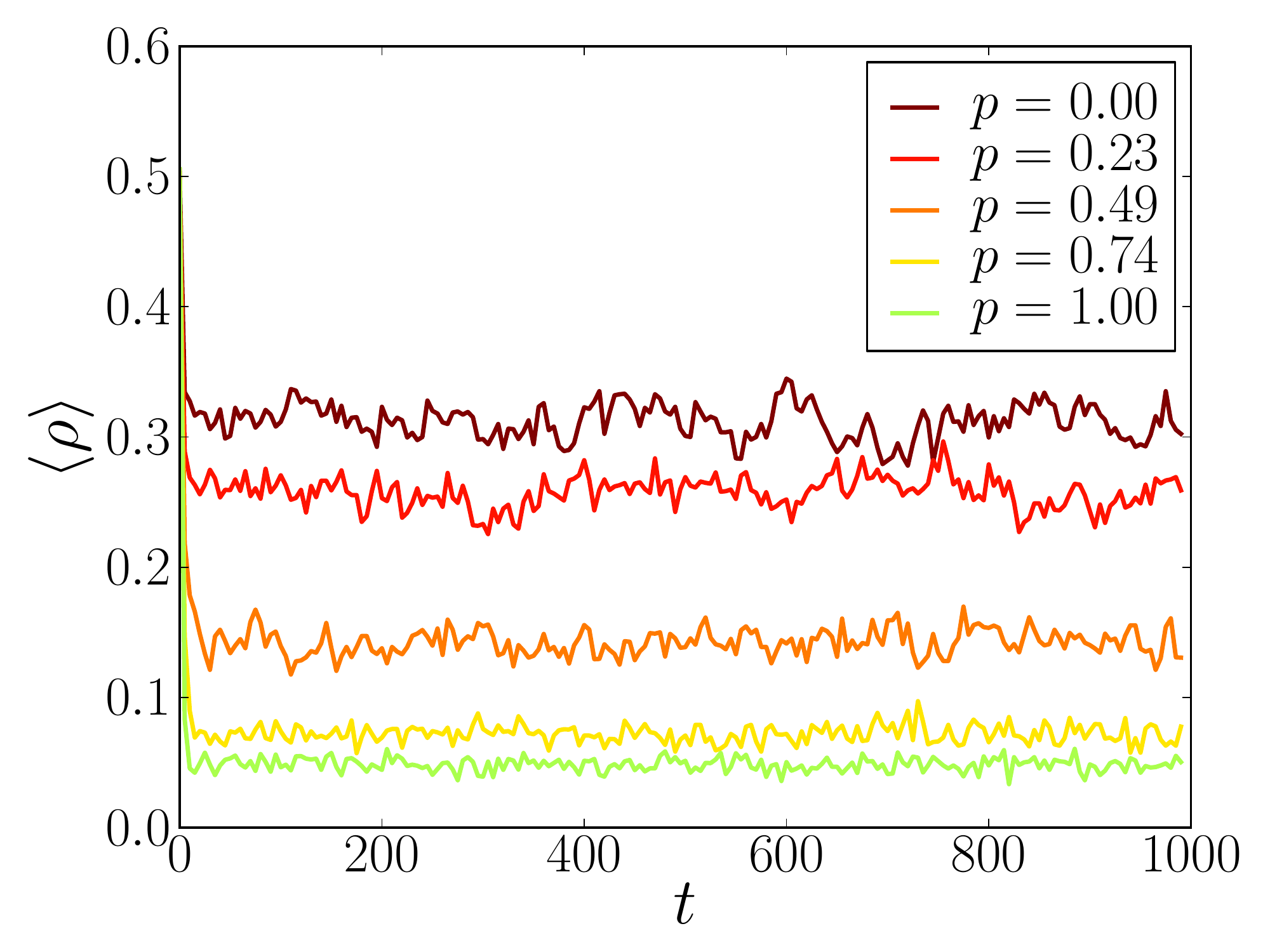}
\label{HomNoise:a}
}
\hspace*{-0.3 in}
\subfigure[\textcolor{black}{$\rho^{\text{asym}}$ from network simulations}]{
\includegraphics[width = 0.35\textwidth]{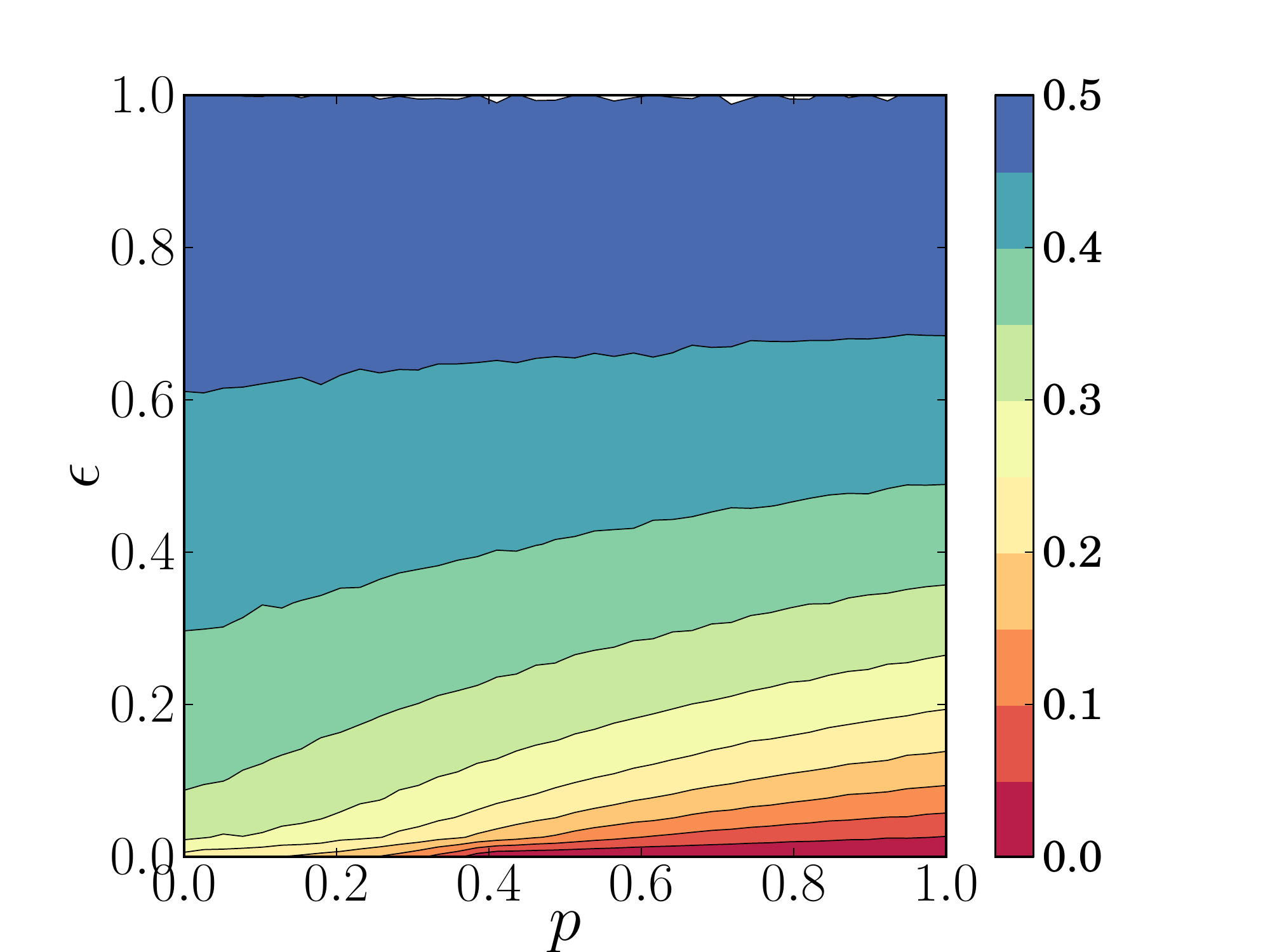}
\label{HomNoise:b}
}
\caption[]{Activity under homogenous noise ($q = 1$), for $N = 250$. \textcolor{black}{\ref{HomNoise:0}-\ref{HomNoise:a}: Measured interface densities averaged over surviving runs in a sample of $10^3$ realizations. The plateaux give the asymptotic values $\rho^{\text{asym}}$. \ref{HomNoise:0}: Activity for noisy ($\epsilon = 0.5$) and noise-free ($\epsilon = 0$) system; \ref{HomNoise:a}: Activity for system with very small noise $\epsilon = 0.03$. \ref{HomNoise:b}: Asymptotic values of average surviving interface density, discretized so that $\epsilon \in [0, 0.03, ..]$.}}
\label{HomNoise}
\end{figure*}
\textcolor{black}{We consider the CVM with homogenous noise, that is, noise that is affecting all ($q = 1$) nodes at some intensity $\epsilon$. The CVM with homogenous noise thus has two parameters, the noise intensity $\epsilon$ and the rewiring  probability $p$.} At $\epsilon = 0$ the model reduces to the original CVM. We compute the asymptotic value $\rho^{\text{asym}}$ of $\left< \rho \right>$, the interface density averaged over active realisations. Homogenous noise destroys the absorbing/fragmentation transition at $p_c$: Fig.~\ref{HomNoise:a} shows \textcolor{black}{that even for a very small noise level ($\epsilon = 0.03$), the computed $\rho^{\text{asym}}$ would be greater than $0$. The explanation is straightforward:
since noise affects all nodes it will eventually destroy any configuration with frozen links, albeit for a very short while. Thus,}  for $\epsilon > 0$ both the absorbing and topologically absorbing states no longer exist, and \emph{any} noise $\epsilon > 0$ is enough to prevent freezing and to keep the system active. It is more active with increasing noise, reaching the maximally disordered phase of $\rho^{\text{asym}} = 1/2$ for $\epsilon = 1$, independent of rewiring (fig. ~\ref{HomNoise:b}).

\begin{figure*}
\hspace*{-0.2 in}
\subfigure[\textcolor{black}{Fixed point of $\rho$ from approximations}]{
\includegraphics[width = 0.4\textwidth, keepaspectratio = true]{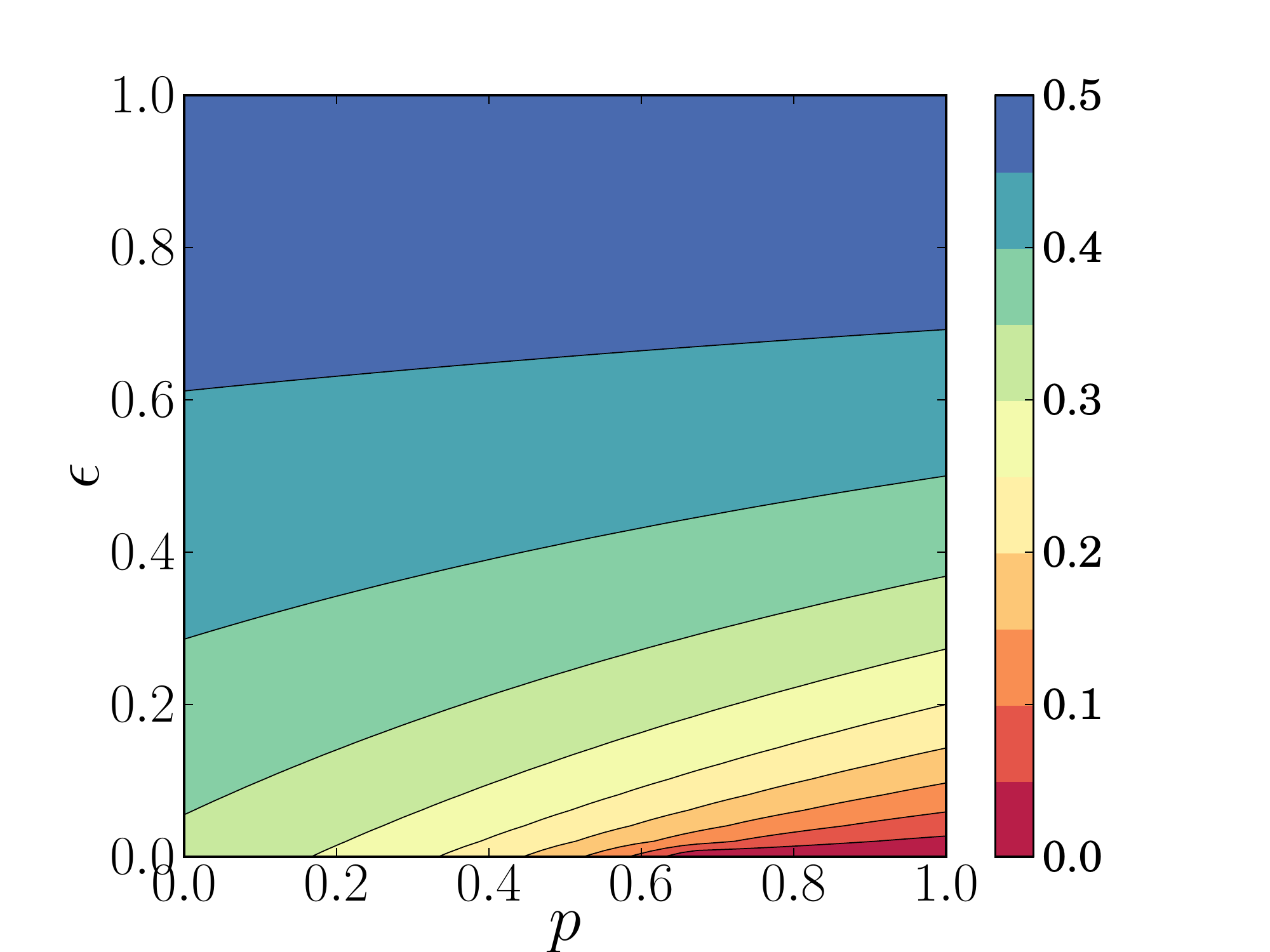}
\label{HomNoiseAn:a}
}
\hspace*{-0.2 in}
\subfigure[\textcolor{black}{Accuracy of analytical approximation}]{
\includegraphics[width = 0.4\textwidth, keepaspectratio = true]{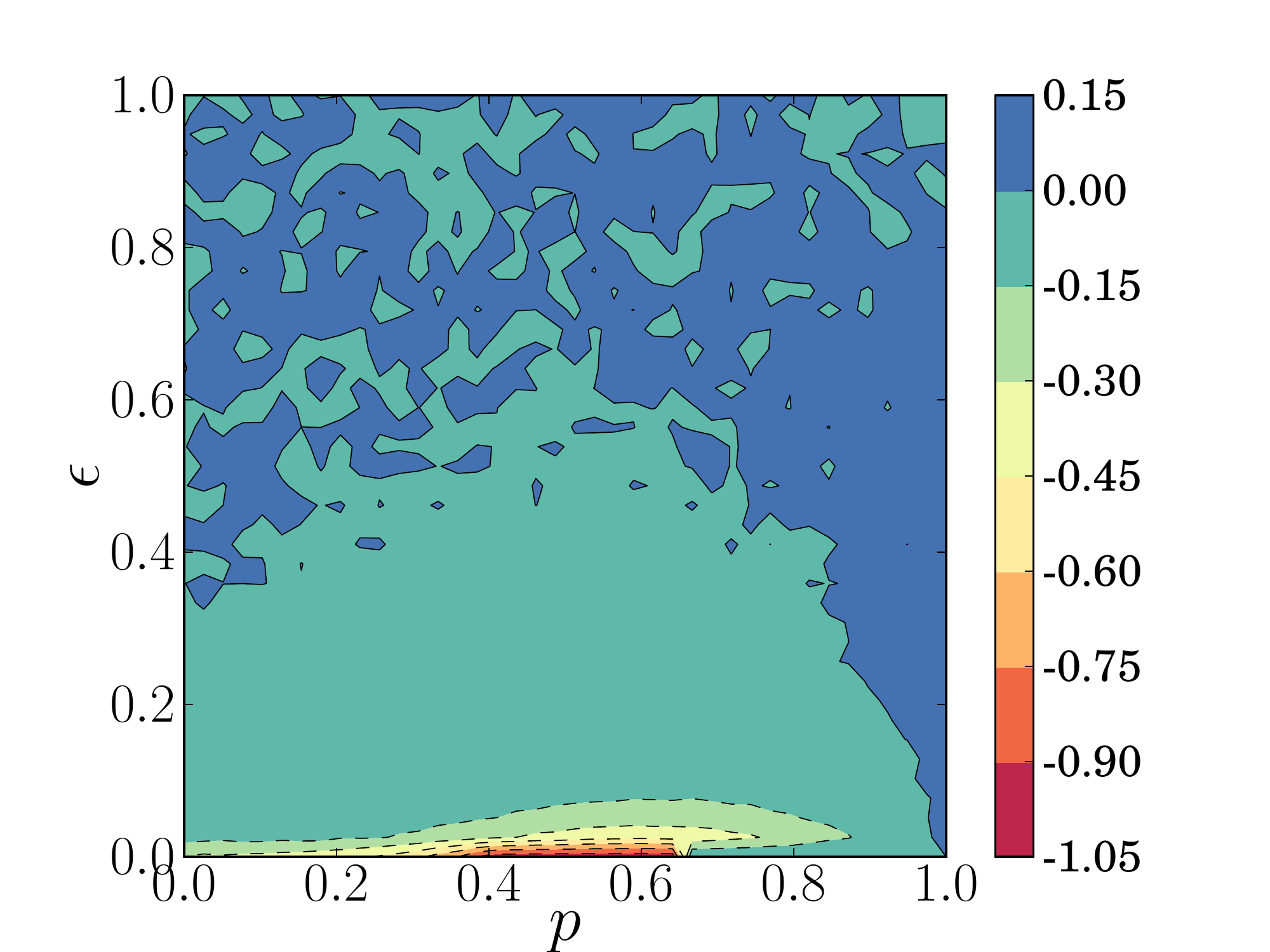}
\label{HomNoiseAn:b}
}
\caption[]{\textcolor{black}{Analytical solution for activity under homogenous noise ($q = 1$). \ref{HomNoiseAn:a}: The steady-state solutions of Eq.~\eqref{noiselast}; \ref{HomNoiseAn:b}: \textcolor{black}{The relative error of the numerical values in fig. \ref{HomNoise:b} w.r.t. the analytical values of fig. \ref{HomNoiseAn:a}.}}
}
\label{HomNoiseAn}
\end{figure*}

This trend in the asymptotic activity can be understood analytically in the thermodynamic limit. We approximate the evolution of the density of active links $\rho$ by adding up the contributions of those updates that would result in a change in $\rho$. This can happen in several ways: 1) selecting an active link and rewiring, followed by a change of state through noise; 2) selecting an active link and rewiring, which is \emph{not} followed by a change of state through noise; 3) changing the state of the node through social imitation with no further action by noise; and 4) changing state of the node \emph{only} as a result of noise, which would be a consequence of selecting a non-active link. Let $\delta_{\text{sc}}$ be the change in the total number of active links given that a node changes state. Then the contributions from the first (1) way of updating is $1 - \delta_{\text{sc}}$, from (2) is $-1$, and from (3) and (4), $-\delta_{\text{sc}}$. Hence we have:
\begin{equation*}
  \Delta \rho = 1/L \sum_k P_k \sum B_{n,k} \left[ \frac{n}{k} \left\{ p(\epsilon - 1)  + (k - 2n) \left(p (\epsilon - 1) + (1-\epsilon/2)\right) \right\}  + \frac{k-n}{k} \left\{ \frac{\epsilon}{2}(k - 2n) \right\}\right],
\end{equation*}
where $L = \mu N/2$ is the total number of links, $P_k$ the probability for a node to have $k$ neighbours, and $B_{n,k}$ the probability of a node with $k$ neighbours to have $n$ of those links being active, making $\delta_{\text{sc}} = k - 2n$. We use a mean-field pair approximation \cite{Vazquez2008,Vazquez2008b,Vazquez2014,Gleeson2013} meaning that the Binomial distribution of $B_{n,k}$ results in $\left<n\right>_{B_{n,k}}=\rho k$, and $\left<n^2\right>_{B_{n,k}}=\rho k +\rho^2k(k-1)$. Since the mean degree $\mu = \left< k \right>$, on rescaling time by $N$ we have
\begin{equation}\label{noiselast}
	\frac{d\rho}{dt} = \frac{4}{\mu} \rho^2 (\mu - 1)(1-p)(\epsilon -1) - \frac{2}{\mu} \rho \left[(2-p)(\mu-1)(\epsilon - 1) + \mu \right] + \epsilon.
\end{equation}
When $\epsilon = 0$ Eq.~\eqref{noiselast} reduces to the noise-free CVM with two stationary solutions: $\rho_1 = 0$ and $\rho_2=\frac{(1-p)(\mu-1)-1}{2(1-p)(\mu-1)}$.
\textcolor{black}{When noise is introduced $\epsilon > 0$, only one stationary solution $0 \le \rho^{*} \le 1$ remains. It is always strictly positive, and is equal to $1/2$ for
$\epsilon=1$ (which means that under full noise the system is fully-disordered). We consider an `active solution' profile as given by a strictly positive solution where it exists,
and by $\rho = 0$ otherwise. This is shown in fig. \ref{HomNoiseAn:a}. Since for noise-free systems $\rho_2$ is an attractor, and for $\epsilon > 0$ only the active solution exists, fig. \ref{HomNoiseAn:b} shows the extent to which the system can be understood through the mean-field node-centric pair approximation. Fig. \ref{HomNoiseAn:b} shows the difference between the numerical and the analytical results, relative to the analytical ones. We choose to show this, rather than the relative error of the approximation, since the zeros of the approximation imply the zeros of the numerical values, but not the other way around. For the majority of the range of $\epsilon$ the error is small, and we conclude that in those regimes eqn. \eqref{noiselast} provides good insight. The visible discrepancy between the two approaches happens for
small noise and $\approx 0.4 < p < \approx 0.7$. We associate the error of around negative unity to the region where the numerical values are zero.
This happens because the numerically computed critical rewiring $p_c \approx 0.38$ (and even less for this finite $N$), whereas $p_c$ given by the analytics is $p^a_c \approx 0.66$, a discrepancy common to this type of analytical approach \cite{Vazquez2008,Gross2008,Gleeson2013,Durrett2012}.
Another curious factor that would result in a discrepancy is because for the noise-free CVM, it is known (\cite{Vazquez2008NJP}) that for $p < p_c$, $\rho^{\text{asym}} = 2/3 \rho_2$, where the prefactor is related to the survival probability. In other words, the analytical solution needs to be rescaled to correspond to the numerical asymptotic value. For small $\epsilon$, taking it into account would further reduce the difference between numerics and analytics; for the majority of the $(p, \epsilon)$ range, on the other hand, the difference is relatively small, so the fixed interface density values derived analytically no longer need to be rescaled using survival probability in order to be comparable to the numerical averages. This is because here survival probability is unity, independent of system size.}

\subsection{Finite-size Effects} 
\label{sub:finite_size_effects}
\begin{figure*}
\hspace*{-0.1 in}
\subfigure[Fully-mixing]{
\includegraphics[width = 0.33\textwidth]{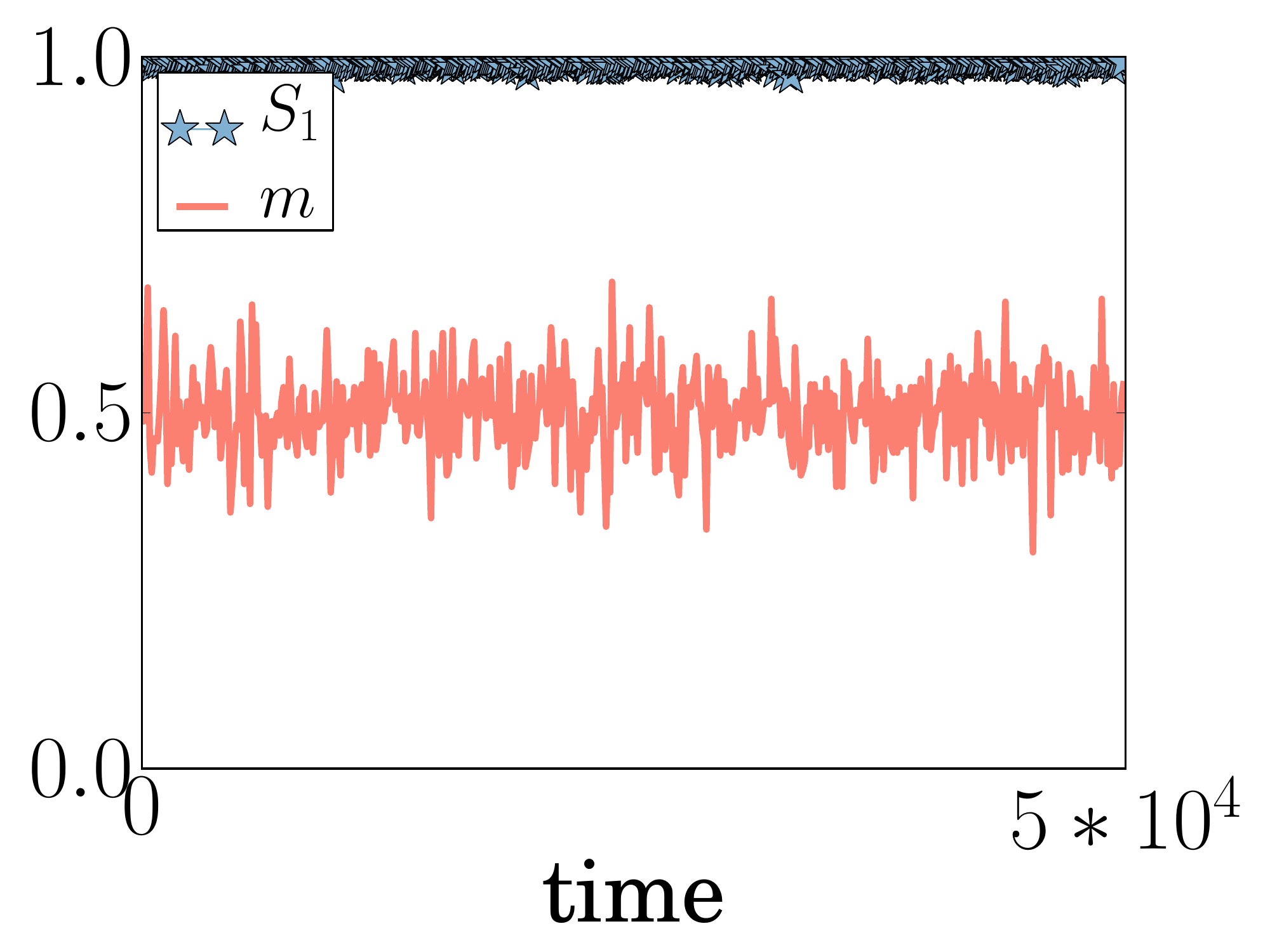}
\label{homnoise_phenomenology:a}
}
\hspace*{-0.2 in}
\subfigure[Bimodal magnetisation]{
\includegraphics[width = 0.33\textwidth]{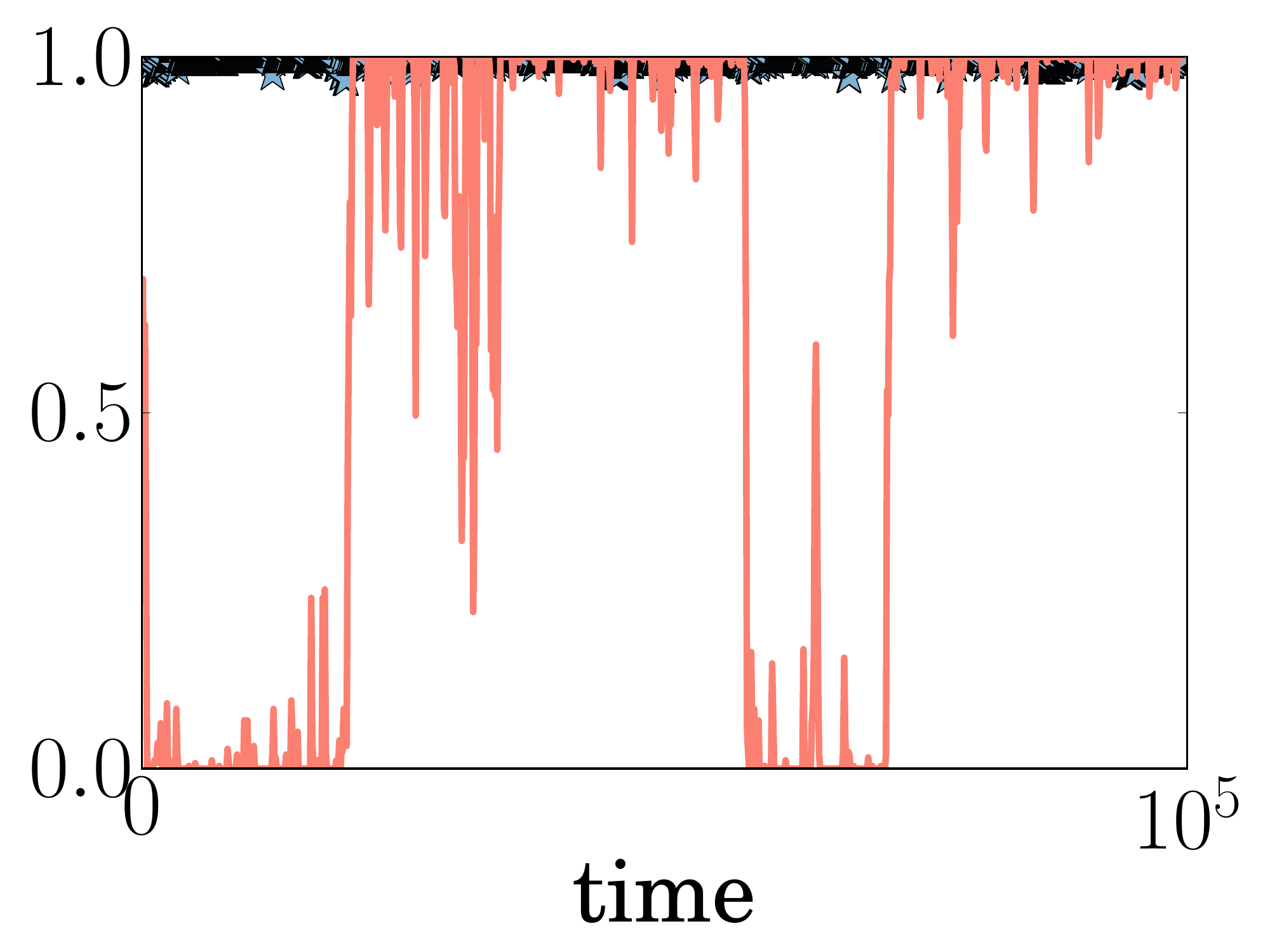}
\label{homnoise_phenomenology:b}
}
\hspace*{-0.2 in}
\subfigure[Dynamic fragmentation]{
\includegraphics[width = 0.33\textwidth, keepaspectratio = true]{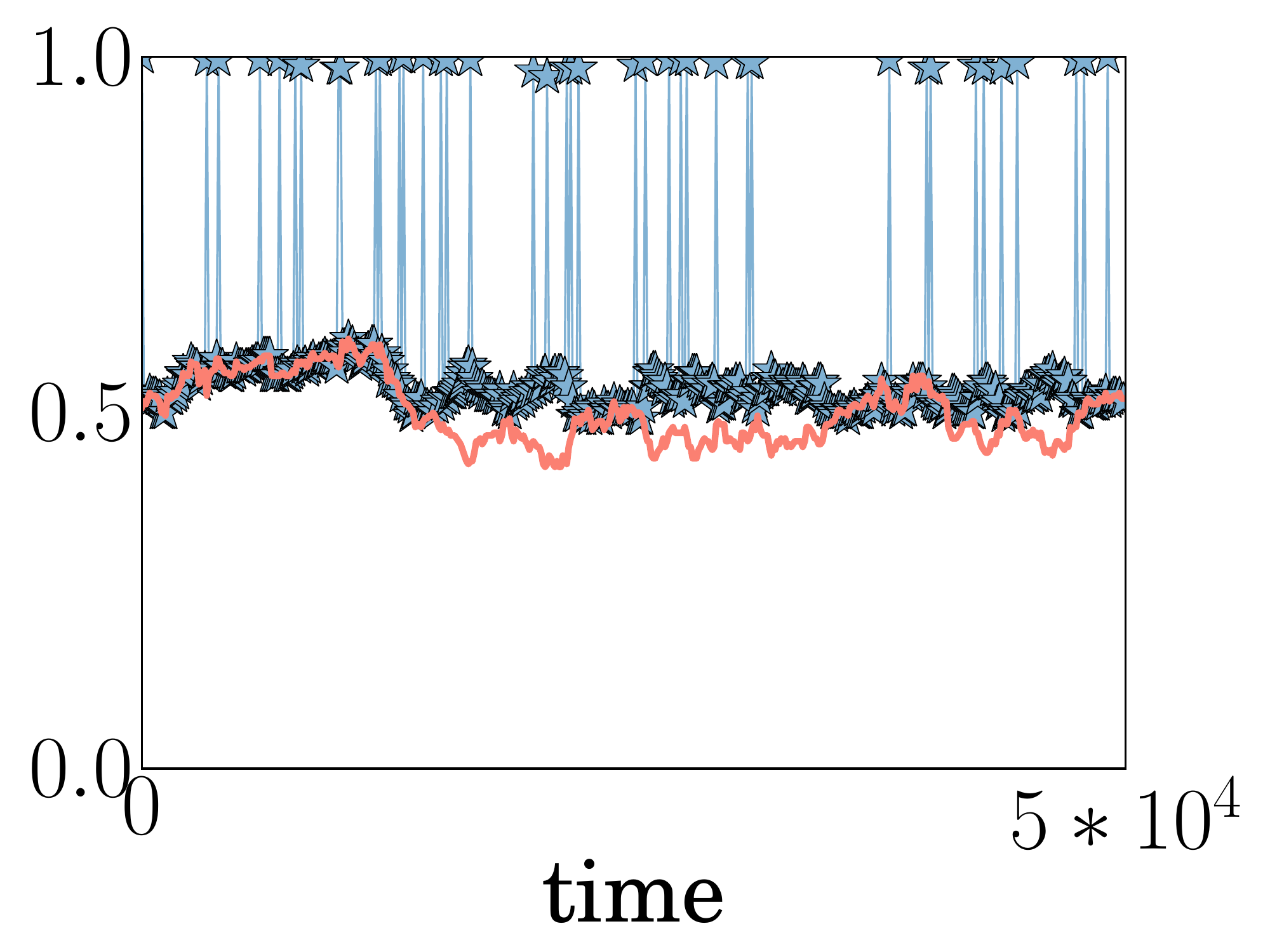}
\label{homnoise_phenomenology:c}
}
\caption[]{Regimes as finite-size effects under homogenous noise. Shown are variation of fraction $m(t)$ (the rescaled magnetisation) of nodes in state $+1$, and the relative size $S_1(t)$ of the largest network component, for a typical time evolution of a single realisation of a system with $N = 250$ nodes. \ref{homnoise_phenomenology:a}: \textbf{The Fully-Mixing} regime characteristic of the thermodynamic limit, demonstrated at $p = 0.5$, $\epsilon = 0.1$. \ref{homnoise_phenomenology:b}: \textbf{The Bimodal Magnetisation}, at $p = 0.2$, $\epsilon = 10^{-4}$. \ref{homnoise_phenomenology:c}: \textbf{The Dynamic Fragmentation} regime, at $p = 0.8$, $\epsilon = 10^{-4}$.}
\label{homnoise_phenomenology}
\end{figure*}

We now consider finite-size phenomenology. We identify three types of behaviour depending on average state of the nodes and the topology of the network. We use the relative number $m$ of nodes in state $+1$ (a rescaled magnetisation) and the relative size of the largest network component $S_1$ to characterize these behaviours. Typical trajectories are shown in Fig.~\ref{homnoise_phenomenology}. At a relatively large noise intensity $\epsilon$, $m(t)$ fluctuates around $0.5$, and $S_1$ approaches $1$ (Fig.~\ref{homnoise_phenomenology:a}). The network stays in one component with occasional isolated nodes, and with approximately equal number of nodes in the two states. We call this the Fully-Mixing regime. When $p < p_c$ and noise intensity is small, $S_1$ continues to be $\sim 1$, but now $m$ switches between trailing the extremes of $0$ and $1$  (Fig.~\ref{homnoise_phenomenology:b}). This corresponds to the network staying in one component and spending long times in a state of consensus of, say, $+1$, and then switching to a state of consensus of $-1$ (the Bimodal Magnetisation regime). Raising $p$ to some $p > p_c$  results in $m$ fluctuating around $0.5$, and $S_1$ abruptly jumping between values around $0.5$ and $1$ (Fig.~\ref{homnoise_phenomenology:c}). In other words, the network is in two giant components, each one in a different state, and these two components continuously recombine and split again (the Dynamic Fragmentation regime).

\begin{figure*}[]
\centering
\hspace*{-0.1 in}
\includegraphics[width = 0.7\textwidth, keepaspectratio = true]{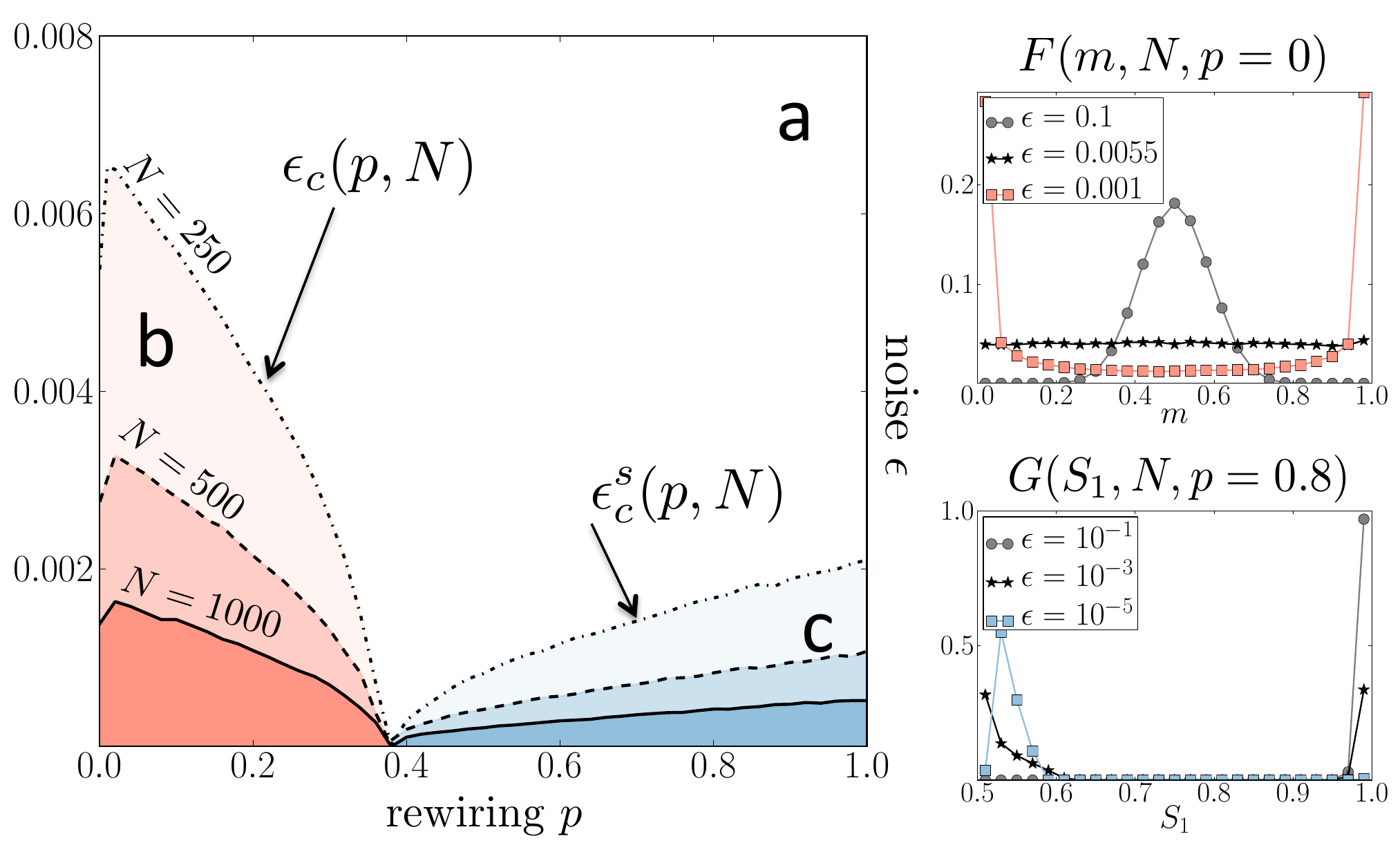}
\caption[]{Phase diagram for finite-size effects under homogenous noise, $q = 1$. The three regions $a$, $b$ and $c$ are defined by the critical noise intensities $\epsilon_c(p,N)$ and $\epsilon^s_c(p,N)$, as well as the critical rewiring probability $p_c \approx 0.38$ of the noise-free CVM. The Bimodal Magnetisation regime (region $b$) exists only for $p < p_c$, and is defined by $\epsilon < \epsilon_c$. The Dynamic Fragmentation regime  (region $c$) exists only for $p > p_c$, and is defined by $\epsilon < \epsilon^s_c$. The Fully-Mixing regime is the complement (region $a$). \textbf{Right Panel}: Three typical trends of the distributions $F(m, \epsilon, N)$ ($p = 0$, top) and $G(S_1, \epsilon, N)$ ($p = 0.8$, bottom; both at $N = 250$) of the (rescaled) magnetisation $m$ and the relative size of the largest component $S_1$, that define the critical noise intensities $\epsilon_c(p,N)$ and $\epsilon^s_c(p,N)$. $\epsilon_c$ is the highest noise giving the weight in the middle of $F(m)$ ($0.25 \le m \le 0.75$) smaller than $0.5$; $\epsilon^s_c$ is the highest noise giving the weight of $G(S_1), S_1 < 0.75$ greater than $0.5$. Both are zero if the sets are empty. Values for $F(m)$ sampled every $100$ time steps from a single network with $N = 250$ evolving until $t = 10^7$, and for $G(S_1)$ until $t = 10^6$, though there are no differences in the trends even if $t$ is varied by an order of magnitude. The measured $\epsilon_c$ is zero for $p \ge p_c \approx 0.38$, and conversely the measured $\epsilon^s_c$ is zero for $p \le p_c$.}
\label{homnoise_main}
\end{figure*}

We summarize these results in the phase diagram shown in Fig.~\ref{homnoise_main}. The Fully-Mixing regime (region $a$) where homogenous noise keeps the network in one giant component with a few occasional solitary nodes, with roughly equal proportion of nodes in differing states, is typical of large noise intensities for finite $N$, or any finite noise intensities in the thermodynamic limit. The other two behaviours are distinct from this based on whether the rewiring probability $p$ is greater than the critical $p_c \approx 0.38$ of the noise-free CVM. Under small noise intensities, $p < p_c$ is characterised by a single component and bimodal magnetisation, whereas for $p > p_c$ there are two giant components in opposing states that continuously split and recombine in the process of dynamic fragmentation (see \cite{videoslink} for animations of the three regimes).

The reason these regions exist is that as $\epsilon$ is lowered, system behaviour approaches that of typical absorbing configurations of the noise-free model, which are qualitatively different to that of the average active state. The two ways they can differ are through $m$ and $S_1$, and these differences manifest themselves respectively for $p < p_c$ and $p > p_c$. If $p < p_c$, the network will freeze a state of consensus, $m = 0$ or $m = 1$, and in one giant component, $S_1 = 1$. For a very small $\epsilon$ systems spend most of their time in these fully-magnetised states, periodically switching between the two extremes of magnetisation, whereas in an active state, $m \approx 0.5$. We therefore define the Bimodal Magnetisation region boundary, $\epsilon_c$, through the change in the nature of the distribution of $m$. An active system does not have an intrinsic preference for one average node state over the other, and therefore as long as there is noise the distribution of $m$, $F(m)$, will be symmetric (here we average over $m(t)$ of a single realisation). This symmetry is broken for $\epsilon = 0$ as the network freezes in one component with either $m = 0$ or $m = 1$. Figure~\ref{homnoise_main} (right panel) demonstrates that as $\epsilon$ is lowered the magnetisation distribution transitions from concave to convex, when the system oscillates between spending time around values associated with the two noise-free absorbing states. We associate the critical $\epsilon = \epsilon_c$ with the flat intermediary stage of the magnetisation distribution, and define the Bimodal Magnetisation regime by $(p < p_c, \epsilon < \epsilon_c)$. \textcolor{black}{In effect, the critical noise $\epsilon_c$ is precisely the noise intensity needed to achieve a balance between the two timescales of the system: the `noise-free' timescale with which the network is driven to consensus, and the timescale on which noise acts to take the system out of it. With too little noise the system oscillates between the two states of consensus, with too much noise there are close to equal number of nodes in each of the states: and the critical noise is the level of external disturbance at which every ratio of nodes in differing states is equiprobable.}

Our simulations indicate that as $\epsilon$ is lowered for $p > p_c$, the giant component begins to occasional split into two halves only to then promptly recombine (Dynamic Fragmentation regime, region $c$). The recombinations are prompted by the appearance of contrarian nodes in among the components characterised by consensus, and the splits happen when rewiring drives the system to the two-component stable state of the noise-free limit. The time spent in two halves increases with lowering $\epsilon$. This dynamic fragmentation transition can be traced through the qualitative change of the distribution $G(S_1)$ of the proportion of time a system spent with relative size of the largest network component being $S_1$. As $\epsilon \rightarrow 0$, $G(S_1)$ transitions from being supported by values of around unity, through having two peaks at $S_1 = 1$ and $S_1 = 0.5$, and until $S_1 \approx 0.5$ at $\epsilon = 0$ (Fig.~\ref{homnoise_main}, right panel). In this context we define $\epsilon^s_c$ as the maximal noise intensity that makes the system spend more than $50\%$ of the time in two components. It is zero for $p < p_c$, and increases for larger $p$, which means that for larger rewiring more noise is needed to keep the system in one component. Hence the Dynamic Fragmentation regime (region $c$) is defined by $(p > p_c, \epsilon < \epsilon^s_c)$.

Similar noise-induced transition \cite{Horsthemke1984} from single to double peaked distribution in magnetisation was observed in an economics model proposed by Kirman in \cite{Kirman1993} (see also \cite{Alfarano2008}), as well as in catalytic reaction models (see \cite{Redner1989} for the mean-field treatment). In the Kirman model agents either change state independently with probability $\epsilon$, or encounter another agent, changing state with probability $1-\delta$ if the new agent is in a different state. A concave/convex transition of the magnetisation distribution occurs at $\epsilon_c$. We therefore see that for $p = 0$ the CVM with homogenous noise is equivalent to the Kirman model, and raising $p$ gives insight into the behaviour of the Kirman model under rewiring. Figure~\ref{homnoise_main} shows that $\epsilon_c$ decreases with $p$ and reaches zero at the critical rewiring $p_c$. Thus, plasticity in network connections shifts downward the critical noise intensity associated with a qualitative change in the magnetisation. Figure~\ref{homnoise_main} also shows that both $\epsilon_c$ and $\epsilon^s_c$ decrease with $N$ and we conclude that neither the bimodal magnetisation nor the dynamic fragmentation transitions exist in the thermodynamic limit.

We use the mean-field treatment to explain the $(p,N)$ dependency of $\epsilon_c$, the onset of the Bimodal Magnetisation regime \textcolor{black}{(see Appendix A for details of the derivation)}. The transition rates show the standard quadratic dependency on $m$, giving after a system size expansion the following Fokker-Planck:
\begin{equation}\label{FP}
	\frac{\partial}{\partial t}P(m,t) = a_1\frac{\partial}{\partial m}\left[(2m-1)P(m,t)\right] + a_2\frac{\partial^2}{\partial m^2}P(m,t) + a_3\frac{\partial^2}{\partial m^2}\left[ m(1-m)P(m,t)\right],
\end{equation}
where $P(m,t)$ is the probability to have a fraction $m$ of nodes in state $+1$ at time $t$, $a_1 = \frac{\mu \epsilon}{2}$, $a_2 = \frac{\mu \epsilon}{4N}$ and $a_3 = \frac{\mu}{N}(1-p)(1-\epsilon)$. Equation~\eqref{FP} is qualitatively similar to the corresponding Fokker-Planck of both the Kirman model and the reaction-desorption system of \cite{Redner1989}. The stationary distribution $P(m)$ of magnetisation undergoes the same bistability transition, and associating $\epsilon_c$, the noise at which $P(m)$ is flat and bistability regime sets in, with the $\epsilon$ at which the derivative $\frac{\partial P(m)}{\partial m}$ at both $m = 0$ and $m = 1$ is zero, we get $\epsilon_c = \left(1 + \frac{N}{2(1-p)} \right)^{-1}$. At $p = 0$ the values of the computed $\epsilon_c$ for different $N$ are in the same order of magnitude as the numerically-obtained ones shown in Fig.~\ref{homnoise_main}. They then monotonically decrease to $0$ at $p = 1$, showing no change at $p_c$ \textcolor{black}{(see fig. 1 in Appendix A)}. For any $p$, as $N \rightarrow \infty$, $\epsilon_c \rightarrow 0$, which means the mean-field analytical approximation supports our numerical results that the bistability regime does not exist in the thermodynamic limit. For small $p$, the agreement of the computed analytical trend with simulations improves with $N$, but necessarily worsens as $p$ increases, since the numerical values drop to $0$ at $p_c$. The discrepancy happens because the absorbing transition cannot be captured by the methodology used above. We therefore conclude that mean-field treatment of magnetisation under the action of noise correctly predicts the behaviour in the thermodynamic limit, and for large but finite $N$ is valid qualitatively only under sufficiently small rewiring.


\section{Targeted Noise} 
\label{sec:targeted_noise}
\begin{figure*}
\hspace*{-0.2 in}
\subfigure[Typical topologically absorbing configurations]{
\includegraphics[width = 0.32\textwidth]{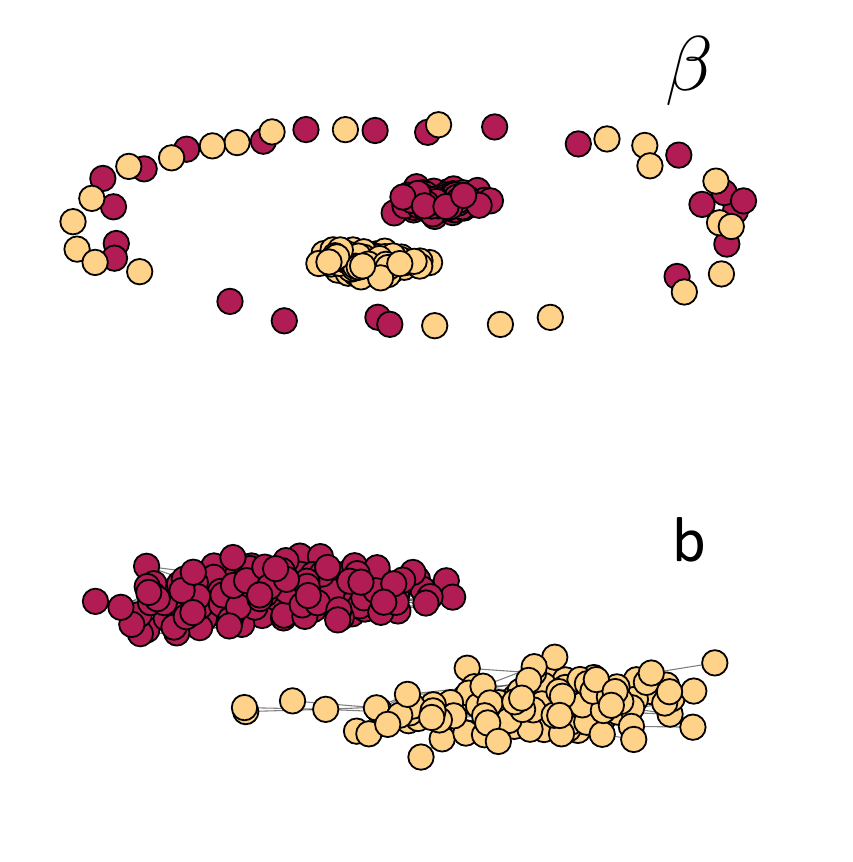}
\label{general:0}
}
\subfigure[Targeted noise fragmentation phase diagram]{
\includegraphics[width = 0.33\textwidth, keepaspectratio = true]{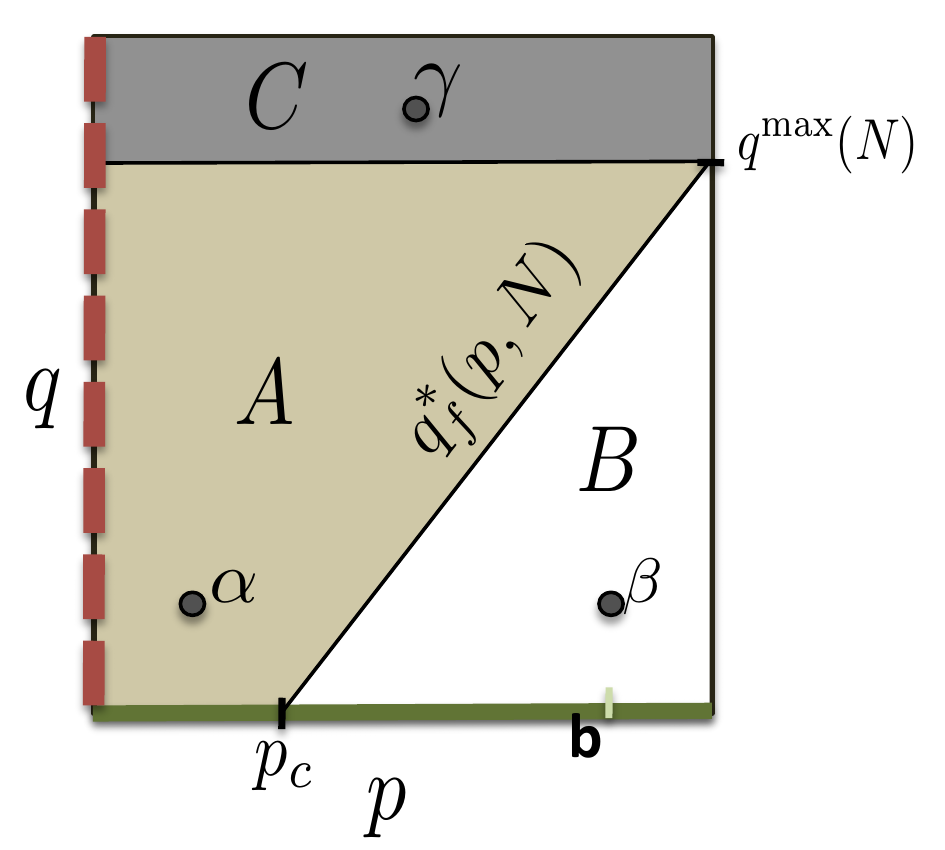}
\label{general:a}
}
\hspace*{-0.2 in}
\subfigure[Network configurations in $\alpha$ at $t = T_{\text{max}}$]{
\includegraphics[width = 0.32\textwidth, keepaspectratio = true]{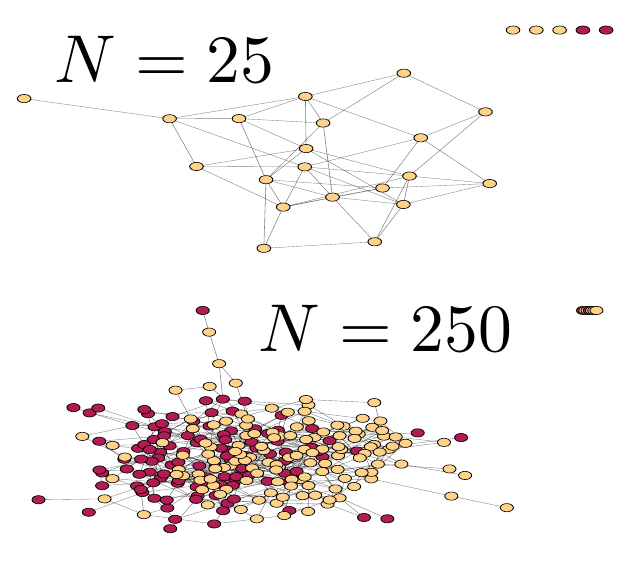}
\label{general:b}
}
\caption[]{\ref{general:0}  Topologically absorbing configuration of $N = 250$ networks at $b:(p = 0.8, q = 0)$ and $\beta:(p = 0.8, q = 0.2)$. Note that those isolated nodes in the topologically absorbing configurations that are the $qN$ noisy nodes will keep changing their state, in both $\alpha$ and $\beta$. \ref{general:a} \textcolor{black}{Schematic representation of the regions} with qualitatively different activity in the (rewiring level $p$, fraction of targeted noisy nodes $q$) space. Thick dark line at $q = 0$ is the Coevolving Voter Model. Thick dashed line ($p = 0$) is the targeted noise equivalent of the Kirman model \cite{Kirman1993}. \textcolor{black}{$q^{\text{max}}$ is the upper limit for the existence of an accessible topologically absorbing state. $q^{*}_f(p,N)$ is the critical noise defining the fragmentation transition.} $\alpha: (p = 0.2, q = 0.2)$, $\gamma: (p = 0.5, q = 0.9)$. \ref{general:b} Fragmentation in region A. At $T_{\text{max}} = 10^5$ the $N = 25$ system shown has reached the topologically absorbing configuration, with one giant component containing nodes all in same state, and $\approx qN = 5$ isolated noisy nodes. The $N = 250$ system is still active at that time, consisting of one giant component with nodes in both states (the spring layout may place some nodes behind others, rendering them invisible), and a few nodes that randomly break away and recombine.}
\label{general}
\end{figure*}

We now confine noise to a targeted subset with relative size $q$ whose nodes change state under noise with probability $1/2$ ($\epsilon = 1$). The two main parameters of the model are therefore $q$ and the rewiring $p$. First we recall the system behaviour in limiting situations. At $q = 0$ the system is the standard CVM, where both the absorbing states and hence the topologically absorbing states exist for all $p$. Conversely, for $p = 0$, the topology is fixed. If $q > 0$ when $p = 0$, the system will have noisy nodes which will not be able to separate from the rest. The random state changes these nodes experience will end up propagating to the rest of the system and keep it active. Therefore no absorbing states will exist here, and in general for any $q > 0$. This is not necessarily the case for the \emph{topologically} absorbing configurations: depending on $q$, networks might reach configurations that will remain constant. A prerequisite for topology to be fixed is for noisy nodes to become isolated. That way their state change will not turn any links into active links, and it is only active links that enact topological change. Therefore for $p > 0$ the topologically absorbing states of the network are characterized by a fragmented network with at least $qN$ components.

Figure~\ref{general:0} demonstrates this fragmentation by showing the difference between the final configurations of a system with no noise (point $b$) and one where a fraction $q = 0.2$ of the nodes are targeted, at some $p = 0.8$ (point $\beta$). Since for no noise this rewiring probability exceeds $p_c$, the network separates into two components. This fragmentation of the network into two components \emph{perseveres} when noise is added, but now it is applicable only to the non-noisy nodes. Thus, apart from $qN$ isolated nodes that keep changing state but continue to be isolated, the remainder of the network is split into two components corresponding to the two available states.

Since the number of links $L$ is conserved, such isolation of $qN$ nodes is only possible if there are enough non-noisy nodes to support all the links. Let $q^{\text{max}}(N)$ be the upper limit for the existence of a topologically absorbing state for the given network ensemble. As $L = \mu N /2$, the non-noisy fraction becomes fully-connected when $(1-q^{\text{max}}(N))N\left((1-q^{\text{max}}(N))N - 1\right) = \mu N$. We neglect corrections of \textcolor{black}{magnitude smaller than and including $(2N)^{-1}$,} and approximate $q^{\text{max}} \approx 1-\sqrt{\frac{\mu}{N}}$. Note that this limit is purely structural and is independent of $p$. We therefore infer that for any finite $N$ there exists a region defined by $q > q^{\text{max}}$ that is characterised by the absence of topologically absorbing states.

We are now in a position to propose a phase diagram in the $(p,q)$ parameter space based on the nature of fragmentation of the final network configurations \textcolor{black}{defined in the limit of $t \rightarrow \infty$}. Based on the observed correspondence of the absorbing and fragmentation transitions in the noise-free CVM, we anticipate this schematic to inform on aspects of activity as well, i.e. how the system behaves in the thermodynamic limit.

Figure~\ref{general:a} shows a sketch of the phase diagram at some finite $N$. Other than the limiting cases of $q = 0$ and $p = 0$ described above, we expect it to have at least two regions. Let point $\beta$ be representative of a set of parameters that lead the network to fragment into the topologically absorbing state defined by \emph{two} giant components in opposite states, and $qN$ isolated nodes, i.e. region B. Let C be the region with $q > q^{\text{max}}$ where the system never reaches the topologically absorbing state but instead remains forever active. C would be present for any finite $N$, and reduce to the $q = 1$ line in the thermodynamic limit. The existence of a third region, region A, is then inferred from numerical simulations, which suggest qualitatively different fragmentation behaviour to the left of a tentative critical targeting line $q^{*}_f(p,N)$. We anticipate A to be the `targeted' equivalent of the `unfragmented' behaviour of the CVM at $p < p_c$, its continuation into the $q > 0$ region: in A we expect networks to freeze in at least $qN$ isolated nodes and \emph{one}, rather than two, giant components. To illustrate this consider a sample point $\alpha \in \text{A}$ shown in Fig.~\ref{general:b}. Just as expected its typical topologically absorbing configuration has $\approx qN$ isolated nodes. The remaining network freezes in one giant component with nodes in the same state, demonstrating the qualitative separation between region A and B.

\begin{figure*}
\hspace*{-0.2 in}
\subfigure[Topological activity]{
\includegraphics[width = 0.33\textwidth]{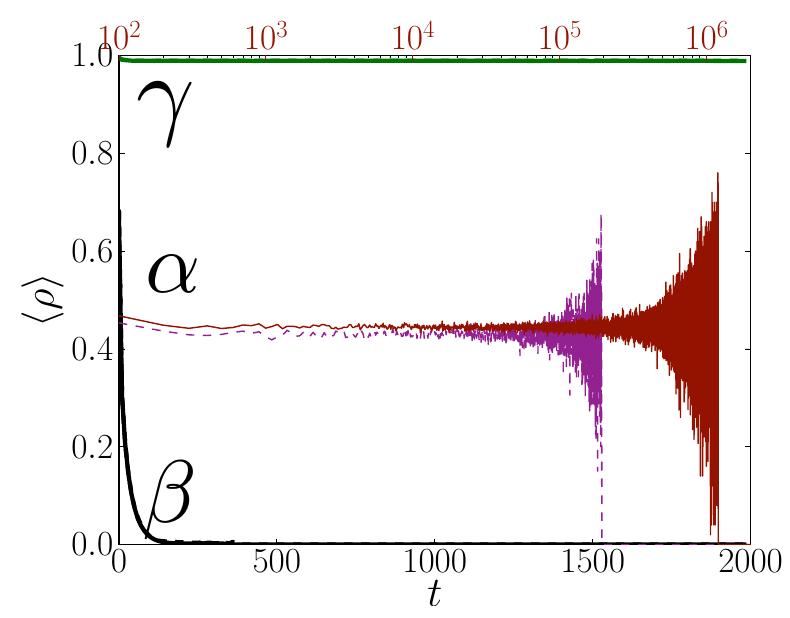}
\label{3PointsActivity:a}
}
\hspace*{-0.1 in}
\subfigure[Relative size of largest component]{
\includegraphics[width = 0.33\textwidth, keepaspectratio = true]{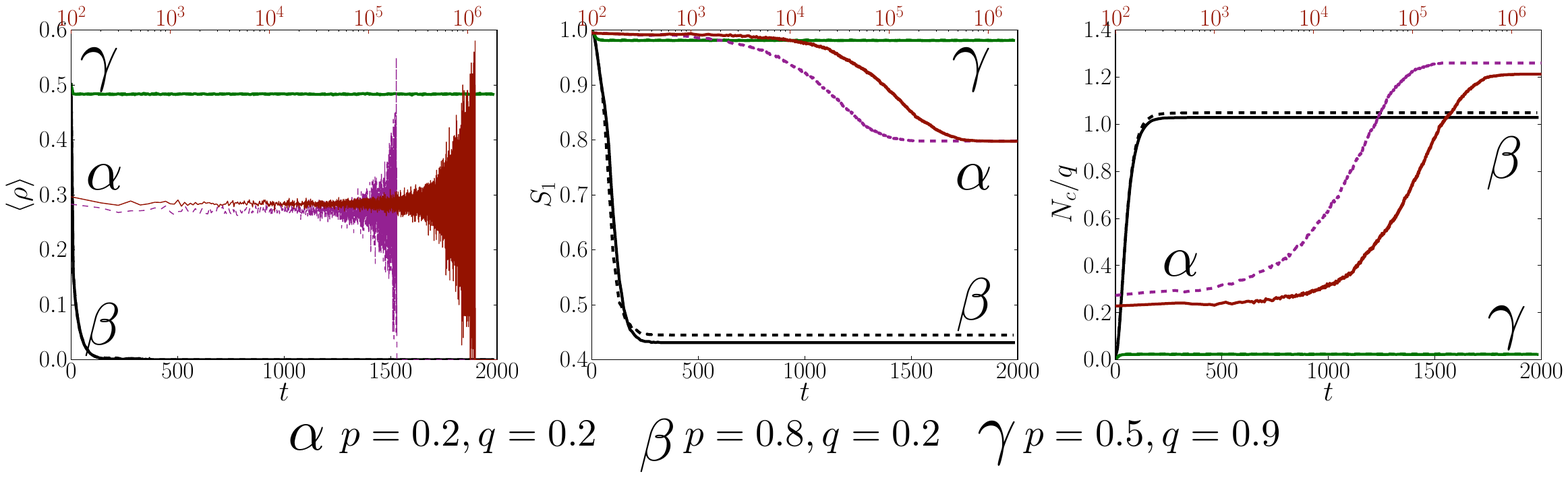}
\label{3PointsActivity:b}
}
\hspace*{-0.1 in}
\subfigure[Relative number of components]{
\includegraphics[width = 0.33\textwidth]{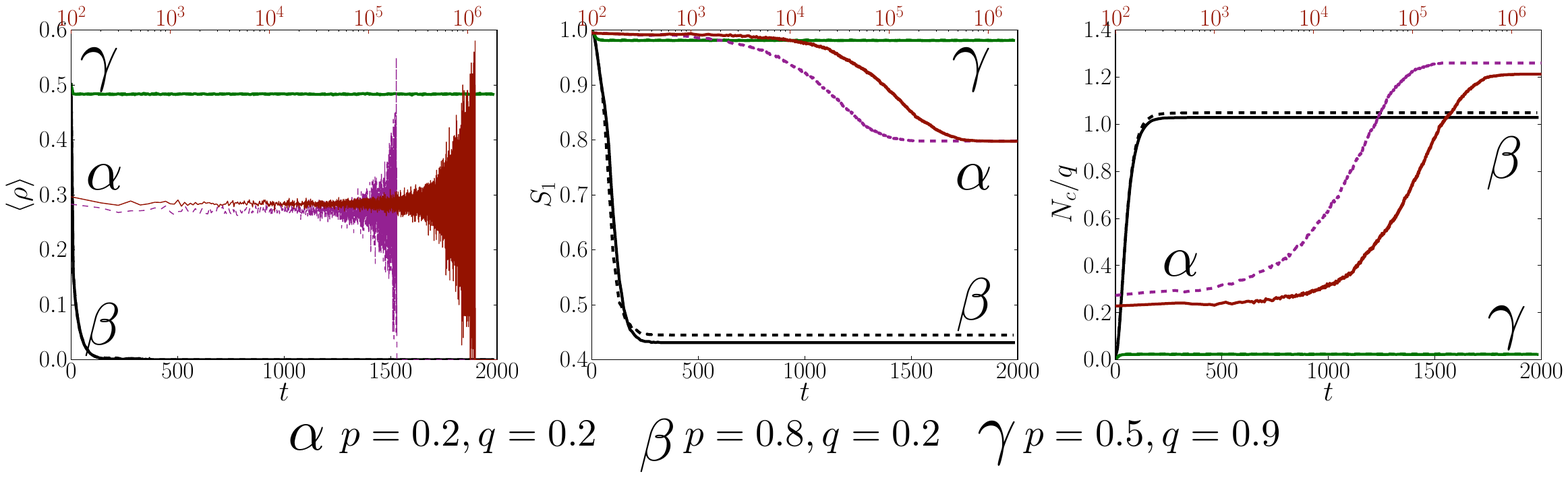}
\label{3PointsActivity:c}
}
\caption[]{\textcolor{black}{Activity patterns with time typical of the three regions in Fig.~\ref{general:a}. \ref{3PointsActivity:a}: Topological activity $\langle \rho \rangle$.} The increase in the $\langle \rho \rangle$ fluctuations is due to the progressively smaller number of surviving runs at larger times ($\langle \rho \rangle = 0$ after the last peak). \ref{3PointsActivity:b}: Relative size $S_1$ of the largest network component, and \ref{3PointsActivity:c}: $N_c/q$, number of components relative to system size ($N_c$) and $q$, both averaged over the complete ensemble. $\alpha$: Upper abscissa, filled line $N = 25$, dashed $N = 20$. $\beta$, $\gamma$: Lower abscissa, filled $N = 500$, dashed $N = 250$. $\alpha$ is at $(p = 0.2, q = 0.2)$, $\beta$ at $(p = 0.8, q = 0.2)$ and $\gamma$ at $(p = 0.5, q = 0.9)$.}
\label{3PointsActivity}
\end{figure*}

The three regions A, B and C can be characterised through the qualitatively different asymptotic behaviour of the ensemble averages of the order parameters of topological activity ($\langle \rho \rangle$) and fragmentation ($S_1$ and $N_c$). Defining $\langle \rho^f_n \rangle$ as the density of frozen links between non-noisy nodes as averaged over surviving runs, we measure the topological activity through $\langle \rho \rangle = 1-\langle \rho^f_n \rangle$. The zeros of $\langle \rho \rangle$ imply that the system is in a topologically absorbing state (whereas the density of active links would not have this correspondence, as now there is a possibility that some frozen links may become active simply because they are joined to a noisy node that may change state). The fragmentation order parameters are the relative size $S_1$ of the largest network component, complemented by the number of connected components relative to system size, $N_c$. Figure~\ref{3PointsActivity} illustrates the variation of these order parameters in time on networks of varying size $N$, computed in the sample points $\alpha$, $\beta$ and $\gamma$ which are taken to be representative of the three regions. Fig.~\ref{3PointsActivity:a} shows that for both $\alpha$ and $\beta$ but not $\gamma$, $\langle \rho \rangle \rightarrow 0$, suggesting that in A and B a finite-size network reaches the topologically absorbing state, while in C it does not. This figure also suggests that in $\beta$ the asymptotic (infinite $N$) $\langle \rho \rangle$ is zero, meaning the system is frozen, while $\alpha$ is an asymptotically active regime for $N \rightarrow \infty$. Thus we associate the border between A and B with an absorbing transition. Our model is different to the CVM in that the active states in A are extremely long lived. The timescale on which the runs in $\alpha$ reach the absorbing state is precisely the reason why in Fig.~\ref{general:0} the typical network configuration in the absorbing state was demonstrated on a network of only $25$ nodes. In fact, \textcolor{black}{in point $\alpha$, increasing the system size from $N = 20$ to $N = 25$ increases the characteristic time $\tau$ to reach the topologically absorbing state from roughly $10^5$ to $10^6$}.

Figures~\ref{3PointsActivity:b} and~\ref{3PointsActivity:c} give topological indicators of the regions.  The active systems in C keep connected in one giant component with $S_1 \approx 1$ (fig.~\ref{3PointsActivity:b}). The marked difference between $S_1$ of $\alpha$ and $\beta$ means that the final configurations in region A will consist of a fraction of $q$ isolated nodes and one giant component, whereas in region B, the non-noisy nodes will instead split into two giant components. This configurational difference echos the fragmentation transition of the CVM, and we propose that in the targeted noise case the fragmentation and absorbing transitions are also coincident. Fig.~\ref{3PointsActivity:c} characterizes this specific nature of the fragmentation by tracing the relative number of components $N_c$. Recall that an absorbing state is one with at least $qN$ isolated nodes. This is what is observed in $\beta$, and what the trend suggest would be observed in $\alpha$ with sufficient network size. (The overshoot at large times for the $\alpha$ curves is due to the system size of networks evolving with parameters given by $\alpha$: here, $N$ is ten times less the size of systems considered at $\beta$. Hence, breakaway nodes at larger times - a consequence of finite size - will mean that $N_c/q$ is visibly larger for relatively small systems. We \textcolor{black}{posit} that for larger system sizes the curves in $\alpha$ will approach $N_c/q = 1$.) The network with parameters given by $\gamma$, however, does not fragment, and has only a small fraction of isolated nodes that constantly combine and split away from the giant component.

\begin{figure}[]
\hspace*{-0.1 in}
\subfigure[Characteristic time $\tau$]{
\includegraphics[width = 0.35\textwidth, keepaspectratio = true]{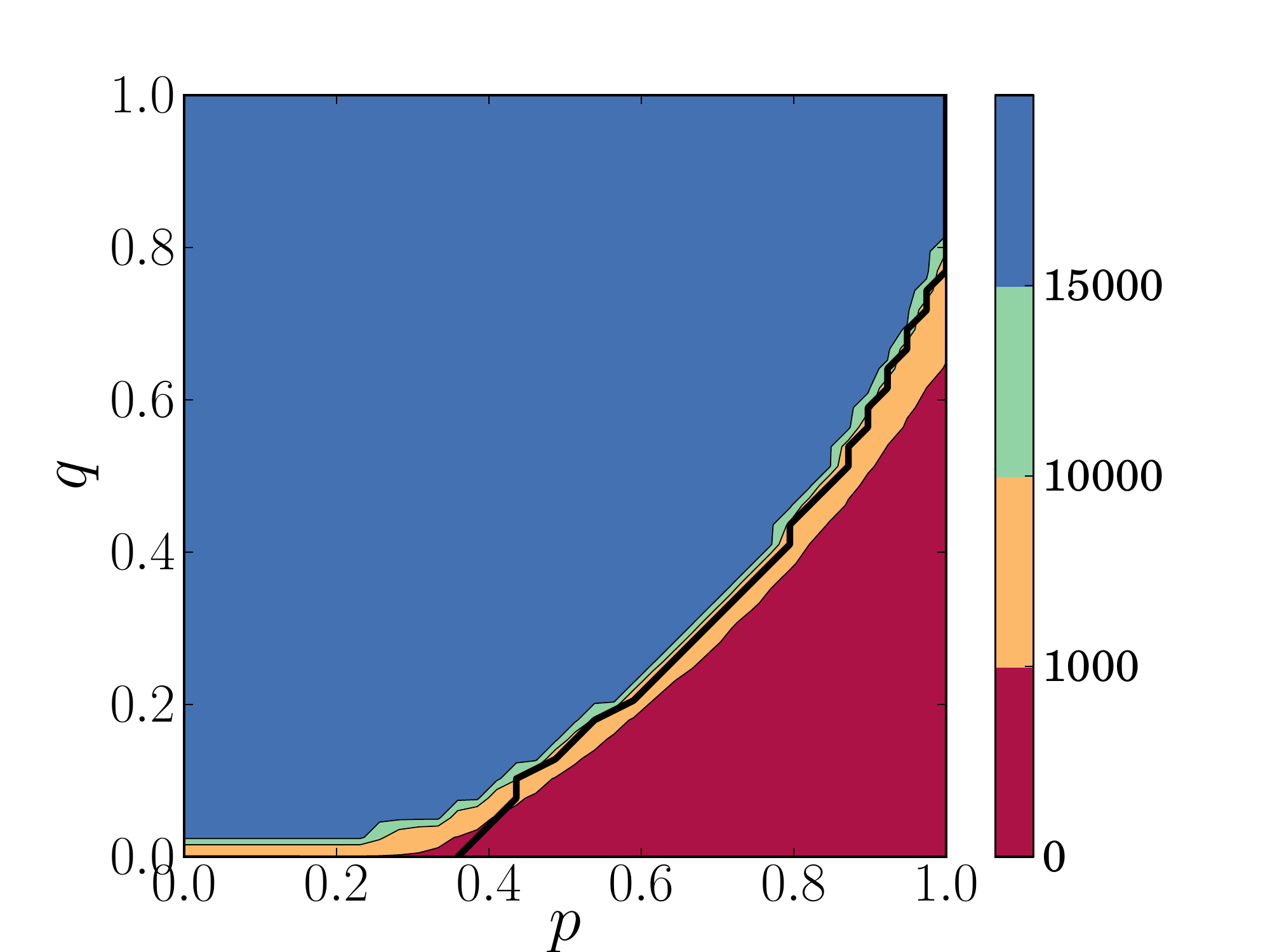}
\label{fragmentation:a}
}
\hspace*{-0.3 in}
\subfigure[Relative size of largest component $S_1$]{
\includegraphics[width = 0.35\textwidth]{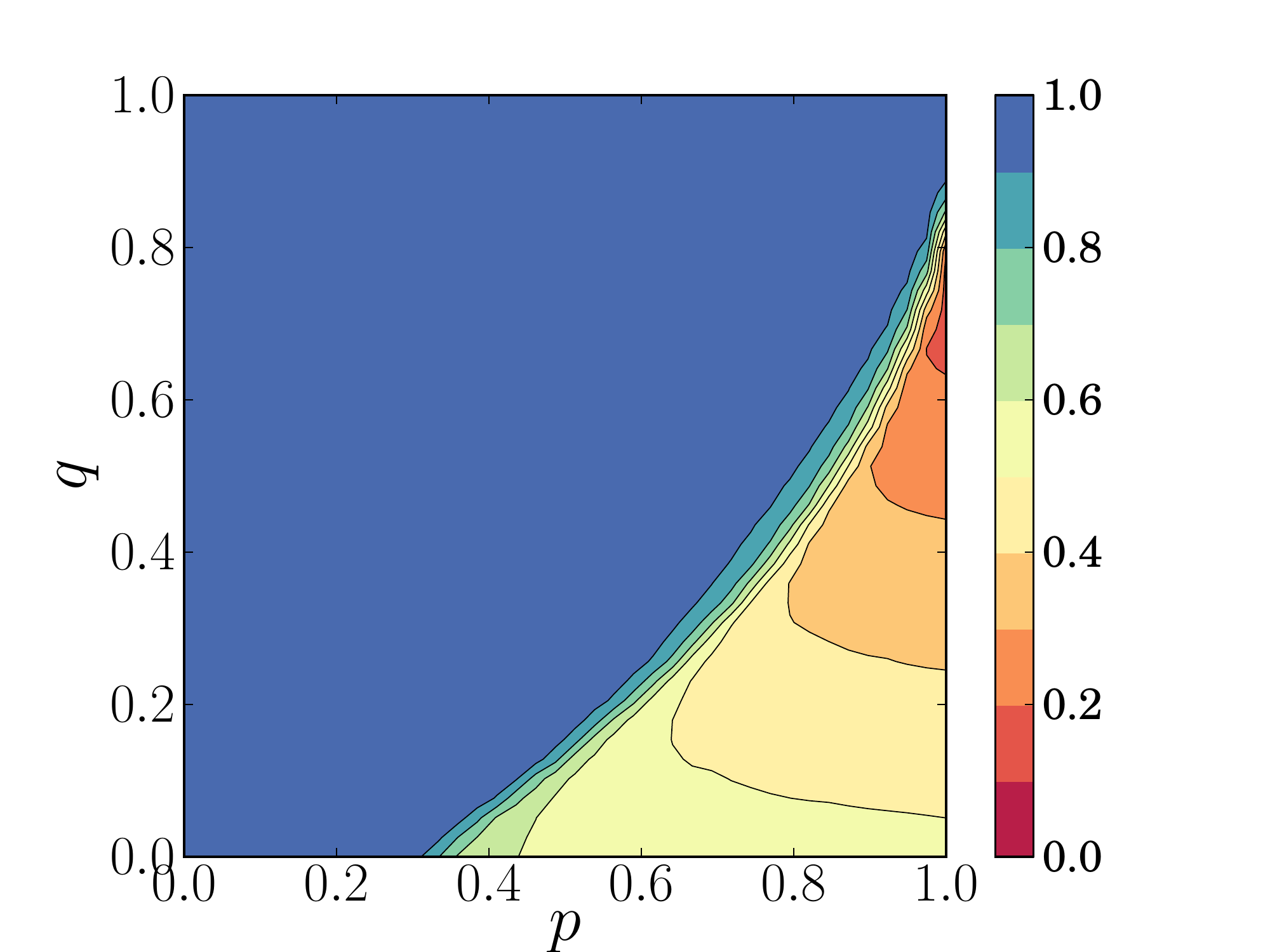}
\label{fragmentation:b}	
}
\hspace*{-0.4 in}
\subfigure[Relative number of components $N_c$]{
\includegraphics[width = 0.35\textwidth]{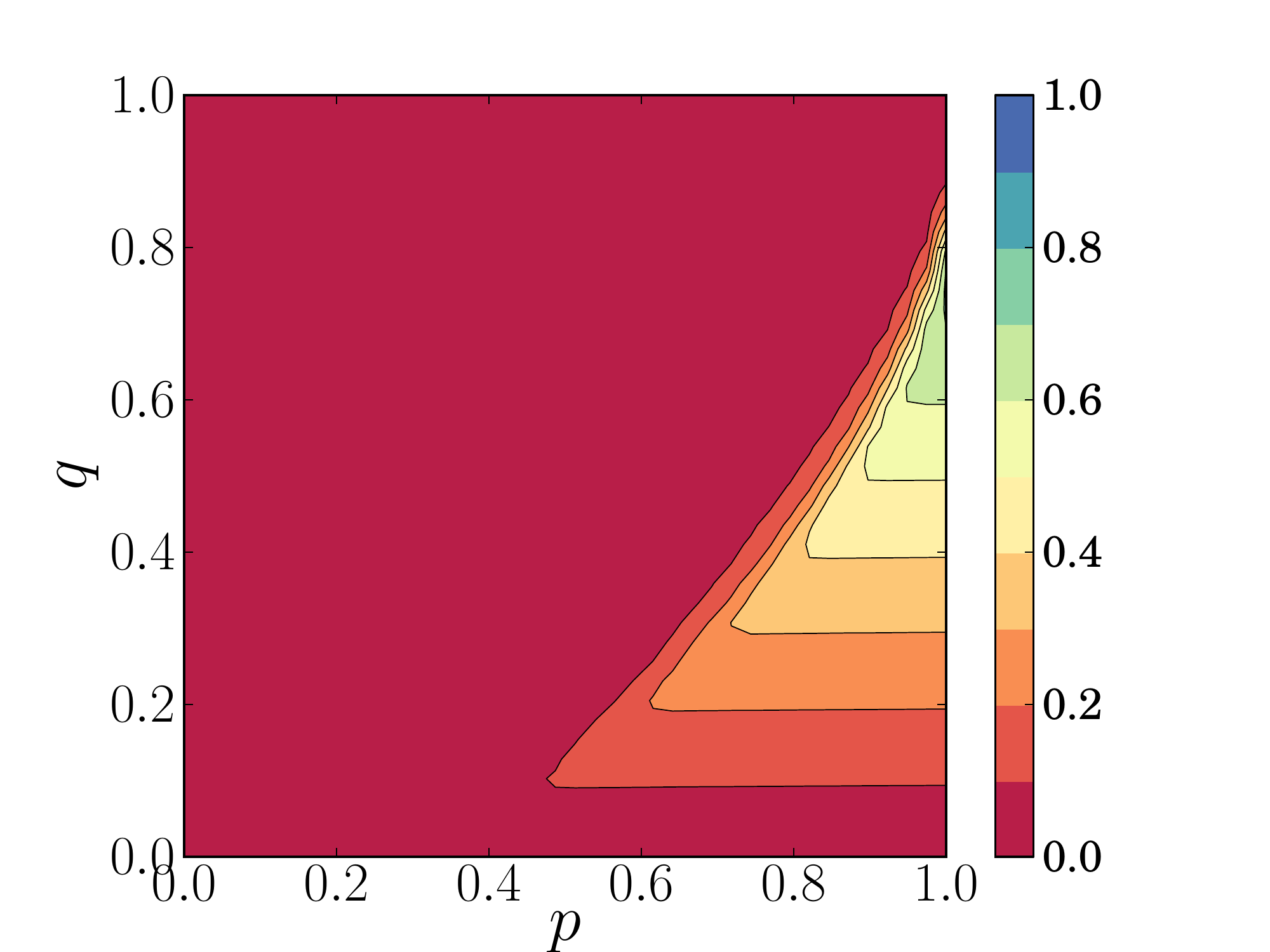}
\label{fragmentation:c}	
}
\caption[]{~\ref{fragmentation:a}: Characteristic time $\tau$, end of scale (blue, color online) corresponding to $\tau$ \emph{upward of} $15000$ timesteps. Thick line: critical $q^{*}_f(p,N, \nu)$ indicative of a fragmentation transition, $\nu = 0.5$.
~\ref{fragmentation:b}-~\ref{fragmentation:c}: Fragmentation profile at some arbitrarily large time $ = T_\text{max}$. Shown are averages over $10^3$ realisations of $N = 250$ network taken at $T_\text{max} = 2000$. ~\ref{fragmentation:a}: Relative size of the largest network component $S_1$.~\ref{fragmentation:b}: Number of components relative to system size, $N_c$.}
\label{fragmentation}
\end{figure}

We now turn to a quantitative characterization of the transition lines between the regions given in fig.~\ref{general:a}. The border between regions A and B is identified with the targeted noise equivalent of the fragmentation transition. We follow \cite{CVM2014} and consider the difference $\Delta S(N)$ between the average sizes of the largest $S_1$ and second-largest $S_2$ components in the topologically absorbing state. $\Delta S(N) = 0$ if the two largest components are equal, and it is around $S_1$ if the size of the second largest component is negligible. We therefore expect region A to be characterized by $\Delta S(N) = 1-q$, and region B by $\Delta S(N) = 0$ \textcolor{black}{(note that because of finite-size effects the values will not necessarily be zero and $1-q$).} Let the critical noise $q_f^*(p,N, \nu)$, defined through the smallest $p$ at which $\Delta S(N) < \nu$, denote the finite-size fragmentation transition (in the thermodynamic limit we expect the transition to be sharper and for the critical noise to become independent of $\nu$). To estimate the critical noise we would need to obtain the values of $\Delta S(N,p,q)$ for the whole range of $p$ and $q$, $q < q^{\text{max}}$. This is a major computational hurdle since, as we have seen before, region A is characterized by long-lived states. Any exploration of the topologically absorbing configurations of reasonably-sized ($> N = 250$) systems is impeded by the superlinear blowup of their characteristic time (see appendix). We notice, however, that the sudden increase of $\tau$ with $q$ is accompanied by a shift in the distributions of $S_1$ and $S_2$ in a manner that reflects $\Delta S(N)$ increasing from zero towards $S_1$. We therefore infer that for supercritical $q$ (i.e. in region A), the topologically absorbing configurations would indeed be characterized by one, rather that two, giant components. We also amend our definition of $q_f^*(p,N, \nu)$ to make it computationally feasible: since we cannot explore all the range of $p$, we associate the sudden increase $\tau$ with the onset of region A, and therefore define $q_f^*(p,N, \nu)$ through smallest $p$ at which $\Delta S(N) < \nu$, \emph{given that the explored $(p,q)$ region has some $\tau(p,q) < \tau_{\text{max}}$}. Increasing the maximal $\tau$ corresponds to exploring further within the A region. That way for any given $\nu$ there is minimum $\tau$ that allows to find $q_f^*(p,N, \nu)$, and a maximal $\tau$ beyond which $q_f^*(p,N, \nu)$ does not change.

Figure~\ref{fragmentation:a} shows the characteristic time, and the computed critical noise $q_f^*(p,N, \nu)$, for $N = 250$, $\nu = 0.5$. The critical noise line denotes the finite-size fragmentation transition as the border between regions A and B. We see that it is coincident with the sudden increase in $\tau$, supporting the notion that the region with long-lived states is identified with the region where the first largest component is much larger than the second largest.

Finally, we characterize the asymptotic behaviour of the system, i.e. the final, topologically absorbing configurations of region B, the long-lived systems in A and the permanently-active systems in C. Figures~\ref{fragmentation:b} and ~\ref{fragmentation:c} show the relative size of the largest network component $S_1$, the relative number of network components $N_c$, as averaged at some large $t$. Since this time limit is larger than the characteristic time for the B region, the values there are averages at the absorbing configurations (and we have checked that the statistics do not change if we instead consider the averages taken at the time the absorbing configurations are reached). They confirm that systems in B freeze the network configuration upon splitting into two giant components and $qN$ isolated nodes (that would keep changing their state). In A and C systems are still active, staying in one component with statistically only a few breakaway nodes that constantly split and recombine. For that system size region C is still non-negligent as $q^{\text{max}}(N = 250) \approx 0.85$, even though here it appears no different from A. Hence, while A and C differ based on the existence of a topologically absorbing state, we see no statistical difference in their active topological configuration. This implies that, since the characteristic time of region A is large and it is unfeasible to define the A-C transition through any potential increase of $\tau$, the border $q^{\text{max}}$ is only relevant for finite-size systems, and in the limit of infinite time.

Figure~\ref{ASID:a} shows the corresponding topological activity phase diagram through plotting the asymptote of the average interface density $\langle \rho \rangle$. For finite $N$ we can define the critical noise $q^*(p,N, \eta)$ through the largest $q$ for which the asymptote of $\langle \rho \rangle$ is within some error $\eta$ to zero. Figure~\ref{ASID:a} shows $q^{*}(p,N, \eta)$ for $\eta = 0.1$. For subcritical noise, networks freeze, staying active otherwise. This activity level decreases with rewiring and increases with the fraction of targeted nodes. We also see that the critical noise trend follows the finite $N$ fragmentation-transition approximation $q^{*}_f(p, N, \nu)$ given in fig.~\ref{fragmentation:a}. Thus numerical simulations lend support to our postulate that the correspondence between the absorbing and fragmentation transition also holds true for targeted noise. The addition of noise, however, shifts the transition to higher values of rewiring.

\subsection{Analytical Approximation for Targeted Noise} 
\label{ssec:analytics}
Standard analytical approximations \cite{Vazquez2008,Vazquez2008b,Vazquez2014} for asymptotic interface density are not suitable to capture the absorbing transition in the targeted noise model, as now the topologically absorbing state is defined not only by the zero of interface density, but also by the placement of the edges. In the topologically absorbing state all links need to be concentrated between `normal' nodes that are not subject to noise. Furthermore, the probability of changing state due to noise is now different between the two node types, which also needs to be incorporated into the model (`noisy' nodes will always be subject to noise, `normal' nodes never). We therefore categorize links into three different types depending on whether or not they join `noisy' nodes, and show that even with the simplest assumptions this approach explains that targeted noise shifts the absorbing transition to higher rewiring probabilities \textcolor{black}{(see the Appendix for detailed derivations)}.

Our model consists of a network of $N$ nodes, $L$ links, and average node degree $\mu$. Each node has a type $n_t$ that for convenience we call `normal', $n$, and `noisy', or random, $r$ (thus $n_t \in \{n,r\}$). This type is assigned to the node at the start and does not change. We will use the variables $x$ and $y$ when referring to node types, so $x$ and $y$ take values $n$ or $r$. This partitions the links into ones that join together two $n$ nodes, two $r$ nodes, and an $n$ and an $r$ node. These three link \emph{types} can be written $xy \in \{nn, nr, rr\}$ ($rn$ is identified with $nr$). Further, since nodes can be in one of two states, we call a link active if it joins nodes currently in different states, and frozen otherwise. There are therefore six \emph{kinds} of links, depending on both the type and states of the nodes at both ends. Let the state vector that gives the density, i.e. the number of such links normalised by $L$, of each link \emph{kind} be $\boldsymbol{\rho}(t) = (\rho_n^a(t), \rho_n^f(t), \rho_r^a(t), \rho_r^f(t), \rho_m^a(t),\rho_ m^f(t))$, where $\rho_n^{a/f}$ is the density of active/frozen $nn$ links, $\rho_r^{a/f}$ the density of $rr$ links, and $\rho_m^{a/f}$ that of `mixed' $nr$ links. The asymptotic behaviour of $\boldsymbol{\rho}(t)$ therefore gives the asymptotic and topologically absorbing states.

The change within each time step to each link kind can be approximated by gathering the respective contributions from the different ways an update can proceed. We identify \textcolor{black}{five} stages to each update. In the first a node $i$ is selected at random. Let $n_t(i) = x$ be its type, which means that $x = n$ with probability $q$, otherwise $x = r$. Next we select a neighbour $j$ at random, $n_t(j) = y$. This means that what is picked is one of the four link kinds potentially joined to that node. This happens with probability $P(xy^{a/f})$, where $y$ can be the same or different to $x$. If an active link is picked then with probability $p$ it is rewired, and with probability $(1-p)$ the node $i$ changes its state. In case of the former the type of new neighbour $k$ will also play a role in determining which link densities are affected. Noise comes in at the final stage where, if $x = r$, then with probability $1/2$ (since $\epsilon = 1$) the state of node $i$ is changed.

In the mean-field limit, given a node of type $x$, the probability $P(xy^{a/f})$ of selecting that link kind is given by the relative number of such links  attached to a node of that type. We use homogenous approximation and do not take into account correlations between link density and node degree. This gives $P(xy^{a/f})$ as just the total number of such links split between the $x$-type nodes. Hence, for $x \neq y$,
\begin{align}\label{probs}
	\begin{split}
	& P(xx^a) = \frac{2\rho_x^a}{N_x^l}  \\ %
	& P(xy^a) = \frac{\rho_m^a}{N_x^l}  \\
	\end{split}
\end{align}
where $N_x^l = 2\rho_x^a + 2\rho_x^f + \rho_m^a + \rho_m^f$ is the relative degree of $x$-type nodes. The probabilities to choose frozen links of these types are given by Eq.~\eqref{probs}, but with $a \rightarrow f$ in the numerator.\\
To approximate the contribution to change in link densities from the different updates we further need $Q(xy^a)$, defined as the change in the density of $xy$ links resulting from an $x$-type node changing state, given the densities $\boldsymbol{\rho}(t)$ directly prior to the state change. For $x \neq y$, these are given by
\begin{align}\label{probs2}
	\begin{split}
	& Q(xy^a) = \frac{\rho_m^a-\rho_m^f}{N_x} \\
	& Q(xx^a) = \frac{2(\rho_x^a-\rho_x^f)}{N_x}, \\
	\end{split}
\end{align}
where $N_n = 2(1-q)/\mu$ and $N_r = 2q/\mu$. Thus for example the probability to select a `random' node, a `normal' neighbour in a different state, to rewire that link to another `normal' node and then to change state due to noise would be written as $qP(rn^a)p\frac{1}{2} (1-q)$, and the resulting contribution to, say, $\rho_m^a$, would be $Q(rn^a)-1/L$.

The contributions from all the processes can be combined to give the following six discrete maps:
\begin{align} \label{allequations}
	\begin{split}
	& \Delta \rho_n^a = 2/\mu \left[  (1-p)(q-1)P(nr^a)Q(nn^a)+(1-p)(q-1)P(nn^a)Q(nn^a)+(q-1)p P(nn^a) \right]\\
	& \Delta \rho_n^f = 2/\mu \left[-(q-1)(P(nr^a)+P(nn^a))((1-q)p+(1-p)Q(nn^a))\right]\\
	& \Delta \rho_r^a = 2/\mu \left[-\frac{q}{2}Q(rr^a)+\frac{q}{2}p(q-1)P(rr^a)+\frac{q^2}{2}p P(rn^a)\right]\\
	& \Delta \rho_r^f = 2/\mu \left[\frac{q}{2}Q(rr^a)+\frac{q}{2}p(q-1)P(rr^a)+\frac{q^2}{2}p P(rn^a)\right]\\
	& \Delta \rho_m^a = 2/\mu \left[-\frac{q}{2}Q(rn^a)-\frac{q^2}{2}p P(rn^a)+\frac{q}{2}p(1-q)P(rr^a)-(1-q)(p+(1-p)Q(nr^a))P(nr^a)-(1-q)(1-p)P(nn^a)Q(nr^a)\right]\\
	& \Delta \rho_m^f = 2/\mu \left[\frac{q}{2}Q(rn^a)-\frac{q^2}{2}p P(rn^a)+\frac{q}{2}p(1-q)P(rr^a) + (1-q)(pq+(1-p)Q(nr^a))(P(nr^a)+P(nn^a))\right]
	\end{split}
\end{align}

\begin{figure}[]
\hspace*{-0.2 in}
\subfigure[Simulation Results]{
\includegraphics[width = 0.45\textwidth, keepaspectratio = true]{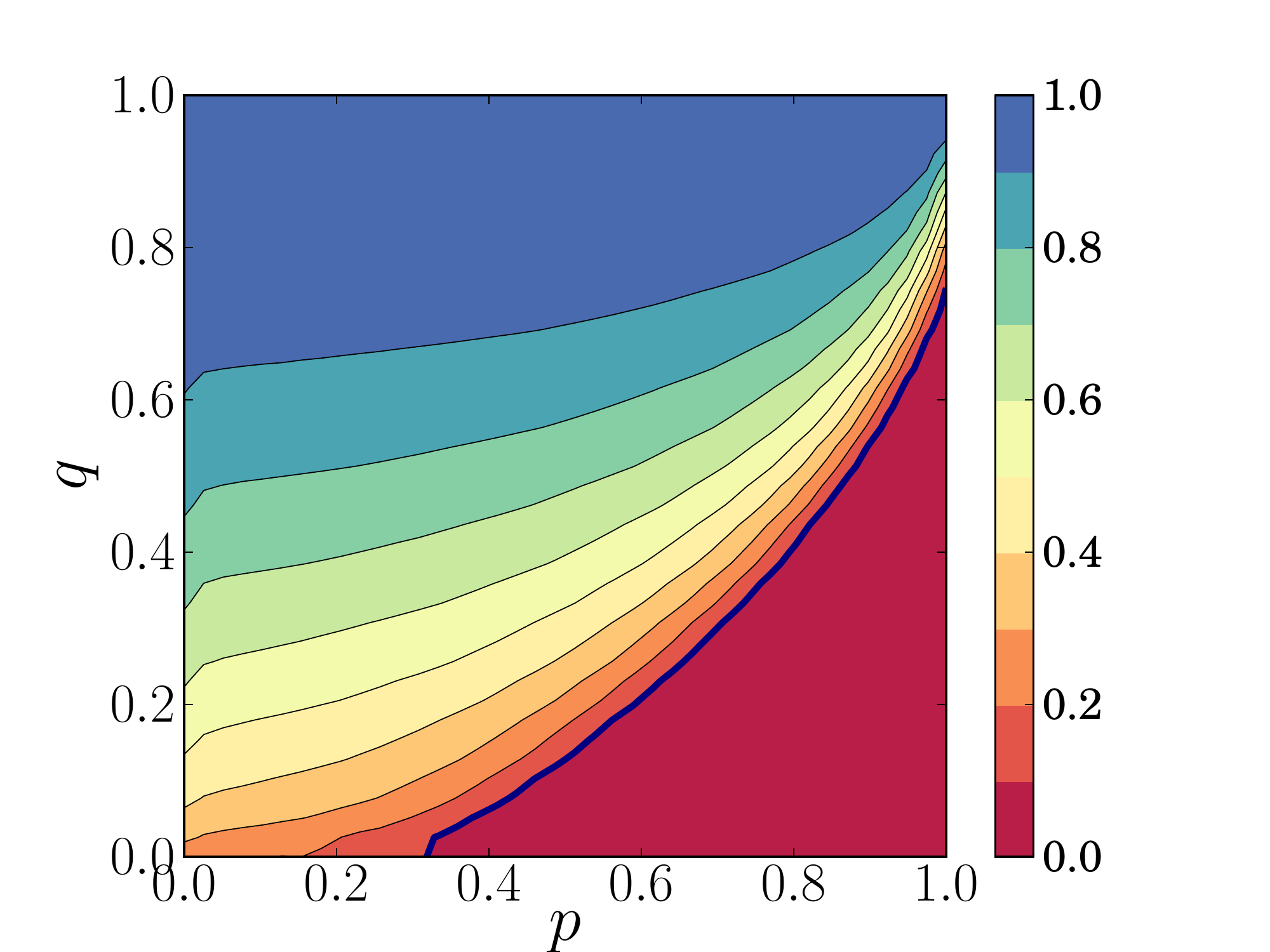}
\label{ASID:a}
}
\hspace*{-0.2 in}
\subfigure[Pair-approximation]{
\includegraphics[width = 0.45\textwidth]{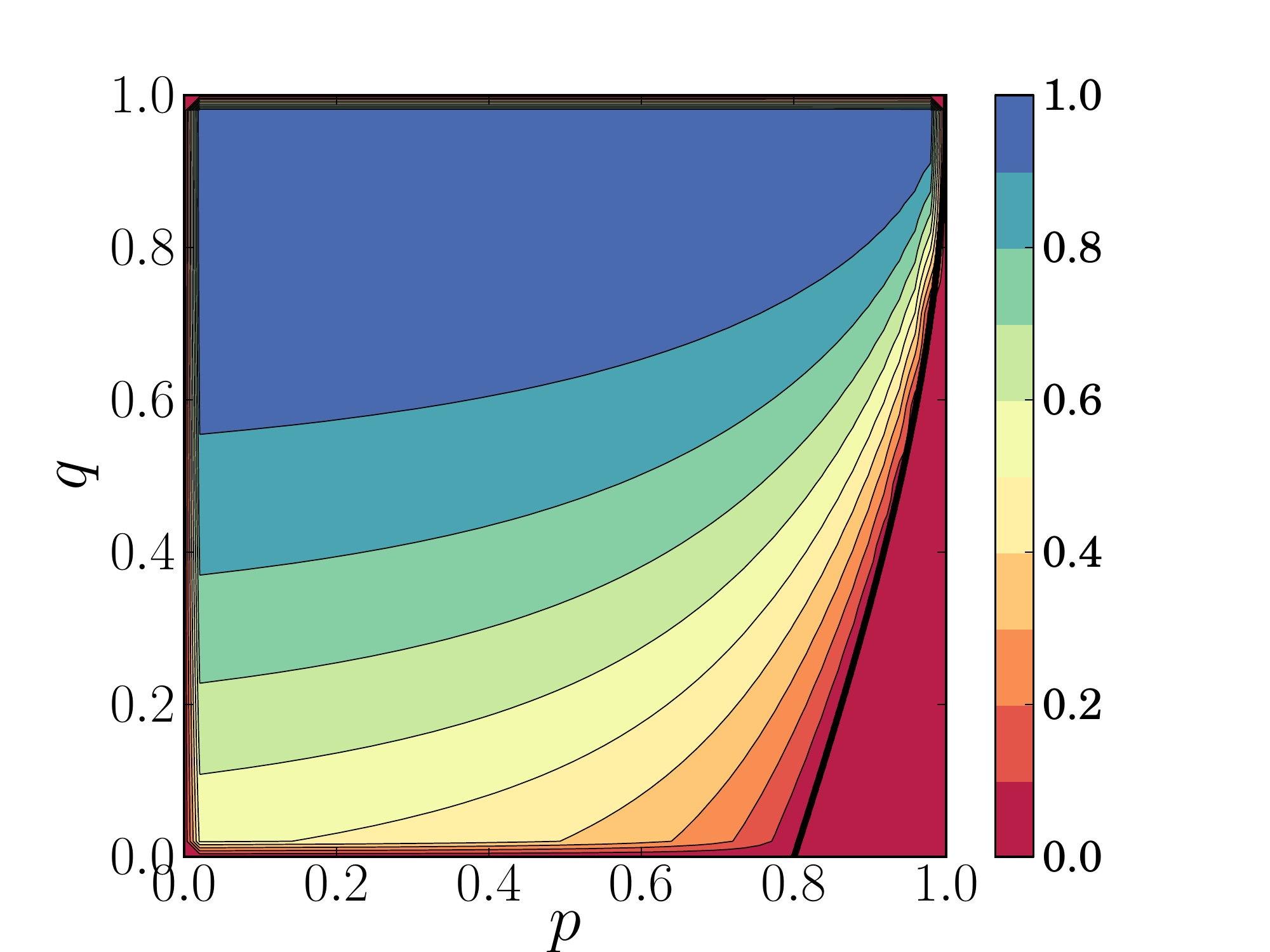} 
\label{ASID:b}	
}
\caption[]{Absorbing transition at $\epsilon = 1$.~\ref{ASID:a}: Asymptotic value of the topological activity $\langle \rho \rangle = 1-\langle\rho^f_n\rangle$, $N = 250$, ensemble size $10^3$, and $\rho^f_n$ is averaged over the surviving realizations. \textcolor{black}{The dark line is the finite-size transition $q^{*}(p, N, \eta)$, $\eta = 0.1$.} ~\ref{ASID:b}: Its analytical approximation obtained as the fixed point solution of Eq.~\eqref{allequations}, for $0 \le p,q \le 1$. The thick black line corresponds to the absorbing transition at the critical targeting $q^{*}(p)$.}
\label{ASID}
\end{figure}

The prefactor of $2/\mu$ comes from normalising a time unit to contain $N$ updates. Since the total number of links $L = \mu N/2$ is conserved, the sum of all equations in system ~\eqref{allequations} is zero. Zeros of ~\eqref{allequations} give the fixed points for $\boldsymbol{\rho}$. $\bar{\boldsymbol{\rho}}_1 = (0, 1, 0, 0, 0, 0)$ is always a solution, though depending on the parameters this could only be accessible through special initial conditions. This solution corresponds to a network that is fragmented in such a way so that all the links are concentrated between `normal' nodes, leaving the `noisy' nodes isolated. This is precisely the type of fragmentation we inferred, and observed in the simulations. Here it is worth noting the main difference between the model and the assumptions behind system ~\eqref{allequations}: the model allows for the saturation of `normal'-`normal' links, which implies the existence of the $C$ region where the topologically absorbing states described by $\bar{\boldsymbol{\rho}}_1$ are unaccessible. In the analytical approximation, however, the node rewires links even if there are no `free' nodes to rewire to, and so here double links (a multigraph) is possible. Hence the mean-field approximations are only valid in the thermodynamic limit, and therefore cannot account for region C (which also does not exist in the thermodynamic limit).

We now consider the limit of either of $p,q$ being zero or unity. The $q = 0$ limit is the standard CVM with no `noisy' nodes. Since now $1 = \rho_n^a + \rho_n^f$, we can use the standard interface density variable $\rho = \rho_n^a = 1-\rho_n^f$. The two solutions are $\rho_1 = 0$ (meaning that $\rho_n^f = 1$, which corresponds to the `frozen' solution $\bar{\boldsymbol{\rho}}_1$), and the active solution $\rho_2 = \frac{1}{2\mu} \frac{\mu(1-p)-p}{1-p}$. The `active' solution starts at $\rho = 1/2$ for $p = 0$ and decreases to zero at $p_c = 4/5$ for $\mu = 4$. We thus see that even the crude, homogenous approximations behind system ~\eqref{allequations} are able to explain the existence of the absorbing transition albeit with a shift in the critical point (approximations in \cite{Vazquez2008} give $p_c \approx 2/3$, whereas numerics have $p_c \approx 0.38$). At $q = 1$, there are no $n$ nodes and the only fixed point is at $\rho_r^a = \rho_r^f = 0.5$, independent of $p$. When $p = 1$, $0 < q < 1$, the only fixed point is $\bar{\boldsymbol{\rho}}_1$. Finally, the case of $p = 0$, $0 < q < 1$ (targeted Kirman model) is a more complicated scenario with fixed planes that is left for future research.

Now consider the general case of $0 < p,q < 1$. Here system ~\eqref{allequations} has one other fixed point (by fixed point we mean a physically relevant one, where all the densities are between zero and unity), which can be computed from the following interrelations
\begin{align}\label{probs3}
	\begin{split}
	& \rho_n^a = \frac{1-q}{2q}\rho_m^a \\
	& \rho_n^f = \frac{(1-q)^2p}{\mu(1-p)} + \frac{1-q}{2q}\rho_m^a \\
	& \rho_r^a = \frac{q}{2(1-q)}\rho_m^a \\
	& \rho_r^f = \rho_r^a\\	
	\end{split}
\end{align}
These come directly from ~\eqref{allequations}. We input these, along with the requirement that the sum is $1$, to express $\rho_m^a$, into the equation for $\Delta \rho_m^f$, to get a quadratic for $\rho_m^a$ (note that these equations are \emph{not} functions of $N$).  At most one other solution is relevant, $\bar{\boldsymbol{\rho}}_2$, which exists for $p < p_c(q)$.\\
Thus, according to our analytical approximations, there are at most two fixed points, $\bar{\boldsymbol{\rho}}_1$ and $\bar{\boldsymbol{\rho}}_2$. We identify the first one with a `frozen' solution where the system has reached a topologically absorbing state. The second fixed point $\bar{\boldsymbol{\rho}}_2$ corresponds to an `active' system where the interface densities plateau. This active solution stops existing at $p = p_c(q)$, beyond which the system is always frozen. Figure \ref{ASID:b} summarizes this asymptotic behaviour by showing the activity level $\bar{\boldsymbol{\rho}}_2$ for $p < p_c$, $\bar{\boldsymbol{\rho}}_1$ for $p > p_c$, and the critical line \textcolor{black}{$p_c$ (identified with the rewiring which realizes the critical targeting $q^{*}(p)$)} which we identify with the absorbing transition. There is very good qualitative agreement with the numerical results (Fig.~\ref{ASID:a}\textcolor{black}{, for a finite $N$}). We see that the simplified pair-approximation correctly predicts both the existence of the absorbing transition, and the monotonic increase of $q^{*}(p)$ to $1$. Moreover, our approximation also explains the `isolated-nodes' aspect of the final network topology, correctly predicting the nature of the observed fragmentation.


\section{Conclusions} 
\label{sec:conclusions_extensions}
We studied two different ways of incorporating noise into a coevolving network model. Homogenous noise that affects all nodes keeps the system topologically active for arbitrarily small noise intensities. Neither the absorbing nor the fragmentation transition of the noise-free system are robust to homogeneous noise. For finite-size systems we distinguish two additional regimes whose appearance can be attributed to the difference between a typical active state in a system with noise, and two distinct frozen states in the subcritical and supercritical noise-free CVM. The first regime of bimodal magnetisation sees the system oscillating between two extremes of consensus. It is bounded by a critical noise intensity $\epsilon_c$ typical of the Kirman model, which here is found to decrease monotonically with rewiring $p$ until reaching zero at the absorbing-fragmentation transition point $p_c \approx 0.38$. The regime of dynamic fragmentation, in which the two halves of the network continuously recombine only to split again, exists for $p > p_c$, and is bounded by $\epsilon^s_c$ which grows with $p$.

We then considered confining noise of full intensity to a fixed subset of the agents of size $q$. We find that targeting the noise in this manner preserves the presence of both the absorbing and fragmentation transitions. These once more coincide and are defined by the critical targeted fraction $q^{*}(p)$. For $q < q^{*}(p)$, systems do not sustain a constant level of topological activity but instead freeze in configurations with two large components in different states, and $qN$ isolated nodes. We identify these isolated nodes as the targeted nodes. As the targeted fraction is raised above the critical value $q^{*}(p)$, the system transitions to a long-lived state with only one giant component and a constant level of topological activity. However, (different) topologically absorbing states still exist, and finite systems will eventually freeze with the `normal' nodes all connected in \emph{one} component, as well as the $qN$ isolated targeted nodes. For finite-size systems we also note the existence of an always-active region, where networks never freeze but sustain a constant level of activity. This region is defined by \emph{over}-targeting ($q> q^{\text{max}}(N)$), which produces a saturation in the connectivity of non-targeted nodes, leading to an overspill of links and consequently an active system.

The critical targeting $q^{*}(p)$ grows monotonically with $p$, and is zero for $p < p_c$. Thus for small rewiring probabilities, $p < p_c$, the existence of long-lived states in targeted noise means that to keep the network topology active it makes no difference whether or not to target the noise. This is not the case for large rewiring. For large $p$ networks freeze under both lack of noise \emph{and} subcritical targeting. When targeting exceeds that threshold, the system is active, just as it would be under arbitrarily minute noise that is spread across all the nodes. As $q^{*}(p)$ grows with $p$, sustaining activity in systems with higher rewiring requires targeting more nodes with noise.

We develop an analytical approximation to compute the densities of links between nodes based on whether these are subject to noise. Our results confirm both the absence of the absorbing-fragmentation transition under homogenous noise, and its presence under targeted noise, as well as the qualitative trend of increase of critical targeting with rewiring.

The fragmentation produced here by targeting noise was first described as \emph{shattered} fragmentation in \cite{CVM2014}, where it was observed to happen on the topologically-fast layer of a multilayer CVM. That setup connected two networks by a fraction of the nodes, and evolved the system layer by layer with a subsequent association of these nodes' states. It is now clear that the topologically-slow layer functioned as an effective noise, and that insight into a multilayer with different timescales can be gained by studying single-layer processes where `noise' represents the state change that is transmitted by the other layer. In fact this insight can work both ways. In \cite{CVM2014} shattered fragmentation occurs for a range rewiring probabilities $p$ of the fast layer. In that model $1-p$ measured how much the fast layer affected the slow layer. We can therefore infer that shattered fragmentation observed here does not require strictly random noise, and that it happens as long as the `random' state change is only weakly correlated with the local neighbourhood.

\textcolor{black}{Future work could consolidate the phenomenology of noise in coevolving networks by assessing the effect of different update types. This work was concerned with vertex-centric update rules, but equivalent edge-centric models can be developed. A preliminary step that is yet missing is understanding whether link dynamics and node dynamics are produce different results in coevolving networks without noise (although in static homogenous networks update types does not make a difference \cite{Suchecki2005}, it could be different if rewiring produces hubs).}

Our results can be a starting point for designing a mechanism for network control. Keeping a coevolving network at some level of activity can not only be achieved by changing the rewiring probabilities, but also by introducing noise. The simplest way would be to target all nodes but should targeting be associated with a cost, our results suggest that depending on the system only a fraction of the population needs to be targeted in order to sustain a constant level of topological activity.\\

This work has been supported by the Spanish MINECO and FEDER under projects INTENSE@COSYP (FIS2012-30634) and MODASS (FIS2011-24785), and by the EU Commission through the project LASAGNE (FP7-ICT-318132).

\appendix
\section{The Fokker-Planck Equation for Magnetization in Coevolving Networks under Homogenous Noise} 
\label{sec:the_fokker_planck_equation_for_homogenous_noise_in_coevolving_networks}
We derive the Fokker-Planck equation for the Coevolving Voter Model under homogenous noise
($q = 1$), with varying noise intensity $\epsilon$. 
Our system consists of $N$ agents, each of which has $\mu$ neighbours. Each agent
can initially be in one of two states, say $+1$ and $-1$, that are assigned independently and randomly. 
The vertex-centric evolution proceeds as follows:
\begin{enumerate}
	\item A node is picked at random. 
	\item If this initial node has any neighbours, then also at random, a 
	neighbour is chosen. Their states are compared, and if different, then one of two things happen:
	with probability $1-p$, the initial node changes its state, becoming the same as the chosen
	neighbour. Otherwise the initial node disconnects from that neighbour and rewires the link to a node that is selected randomly
	from the set of those nodes that are both as yet disconnected to the initial node, and are
	in the same state as it. If no such node exists, no rewiring takes place, and two nodes
	stay connected.
	\item With probability $\epsilon/2$ the initial node changes state.
\end{enumerate}
Note that the final step, which simulates the action of noise independent to the system, happens
regardless of any changes to the network or its states that may or may not have taken place in the previous step.\\ 

We now write the mean-field transition rates for the change in the number $n$ of agents in state $1$. Let $m = n/N$. Then
\begin{align}\label{probs3}
	\begin{split}
	P(m \rightarrow m + 1/N) & = m(1-m)\mu A + (1-m)^2 \mu B \\
	P(m \rightarrow m - 1/N) & = m^2\mu B + m(1-m) \mu A,
	\end{split}
\end{align}
where $A = (1-p)(1-\epsilon/2)+p\epsilon/2$, and $B = \epsilon/2$. \textcolor{black}{The Fokker-Planck equation
associated with the master equation is:}
\begin{equation}\label{FP}
	\frac{\partial}{\partial t}P(m,t) = a_1\frac{\partial}{\partial m}\left[(2m-1)P(m,t)\right] + a_2\frac{\partial^2}{\partial m^2}P(m,t) + a_3\frac{\partial^2}{\partial m^2}\left[ m(1-m)P(m,t)\right],
\end{equation}
where $a_1 = \frac{\mu \epsilon}{2}$, $a_2 = \frac{\mu \epsilon}{4N}$, $a_3 = \frac{\mu}{N}(1-p)(1-\epsilon)$, \textcolor{black}{and time has been rescaled by $1/N$}. Written in this way, for $p = 0$ the Fokker-Planck becomes comparable to that found in \cite{Redner1989}, where the first two terms arise from the desorption, and the last term from the reaction processes. The model in \cite{Redner1989} had an effective noise on the links; the similarity of the mean-field treatment suggests that noise on the links can be remapped to noise on the nodes. We also note that all three terms will be present in the Fokker-Planck written for the Kirman model \cite{Kirman1993,Alfarano2008}. The difference between these models and the one we are introducing is in the relative weight of the three terms.\\
The relevant stationary solution to eqn.~\ref{FP} is 
\begin{equation}\label{Px}
	P(m) = P(0)\left[ 1-\left( \frac{2m-1}{m_0} \right)^2 \right]^{- \alpha},
\end{equation}
where $m_0^2 = \frac{4a_2 + a_3}{a_3}$, $\alpha = 1 - \frac{a_1}{a_3}$, and $P(0)$ the normalizing constant. Substituting, we get 
\begin{equation*}
	x_0^2 = 1 + \frac{\epsilon}{(1-p)(1-\epsilon)}, \, \, \alpha = 1 - \frac{\epsilon N}{2(1-p)(1-\epsilon)}.
\end{equation*}
The stationary distribution of magnetization given by eqn.~\ref{Px} undergoes a bistability transition, similar to the one observed numerically (see main body of the paper for figures). We associate the $\epsilon_c$, the noise at which $P(m)$ is flat and bistability regime sets in, with the $\epsilon$ at which $\frac{\partial P(m)}{\partial x} \bigg|_{m = 1} = 0$. This happens when $\alpha = 0$, giving $\epsilon_c$ as
\begin{equation}\label{ec}
	\epsilon_c = \left(1 + \frac{N}{2(1-p)} \right)^{-1}. 
\end{equation}

\begin{figure*}
\hspace*{-0.2 in}
\includegraphics[width = 0.5\textwidth]{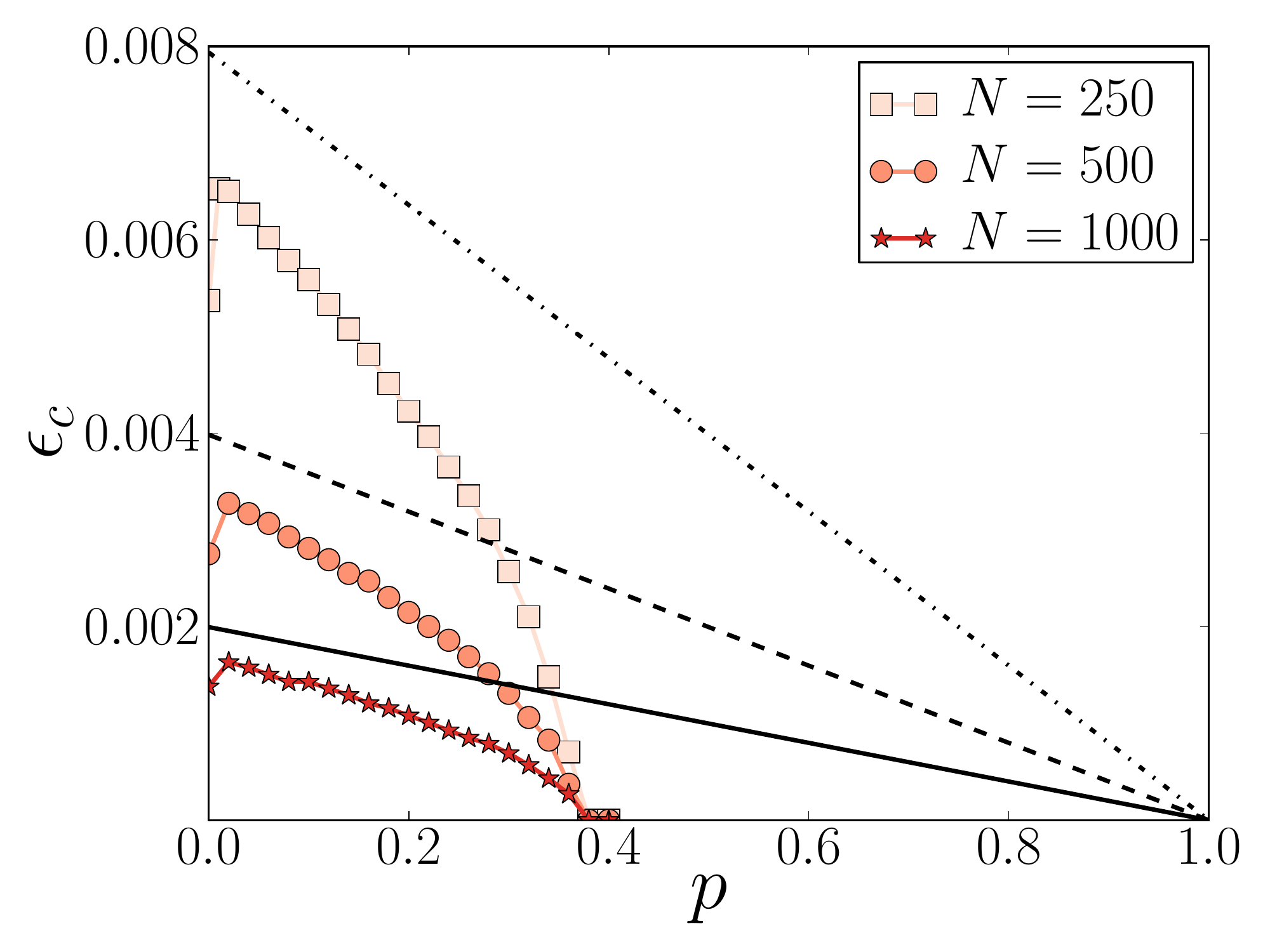}
\caption[]{Critical $\epsilon_c$ for homogenous noise. Plots with markers are the numerical results shown 
in the main body of the paper. The point at which these become zero corresponds to the critical rewiring $p_c$. Straight lines are analytical estimates given by eq.~\eqref{ec}, for the three respective values of $N$ (traversing top to bottom increases $N$).}
\label{FP}
\end{figure*}

Figure~\ref{FP} compares the numerical and the analytical results for the critical noise. At $p = 0$ the values of $\epsilon_c$ computed from eqn.~\eqref{ec} for different $N$ are of the same order of magnitude as the ones obtained numerically. The analytical values then monotonically decrease to $0$ at $p = 1$, showing no change at $p_c$. For any $p$, as $N \rightarrow \infty$, $\epsilon_c \rightarrow 0$, which means the bistability regime does not exist in the TD limit. This corresponds to what is observed in simulations. For small $p$, the agreement of the computed analytical trend with simulations improves with $N$, but necessarily worsens as $p$ increases, since the numerical values drop to $0$ at $p_c$. This happens because the absorbing transition cannot be captured by the methodology used above. We therefore conclude that mean-field treatment of magnetisation under the action of noise is only valid under sufficiently small rewiring.

\section{A-region Statistics} 
\label{sec:b_characteristic_time_and_other_statistics_}
Region A is characterized by long-lived states. We have seen 
that a system with $N = 20$ nodes can take up to $10^5$ timesteps
to reach a topologically absorbing state, and for a system with $N = 25$ nodes 
this time increases one order of magnitude. It becomes computationally unfeasible to 
wait until the entire ensemble of reasonably-sized system (e.g. with $N = 250$) reaches 
the topologically absorbing states, least of all to do for a wide
range of parameters. Instead, we explore the A region by venturing
deeper into it starting from the borders at small and large $p$, 
and noting the changes on $\tau$ and other statistical
characteristics.

\begin{figure*}
\hspace*{-0.2 in}
\subfigure[$\tau/N$]{
\includegraphics[width = 0.3\textwidth, keepaspectratio = true]{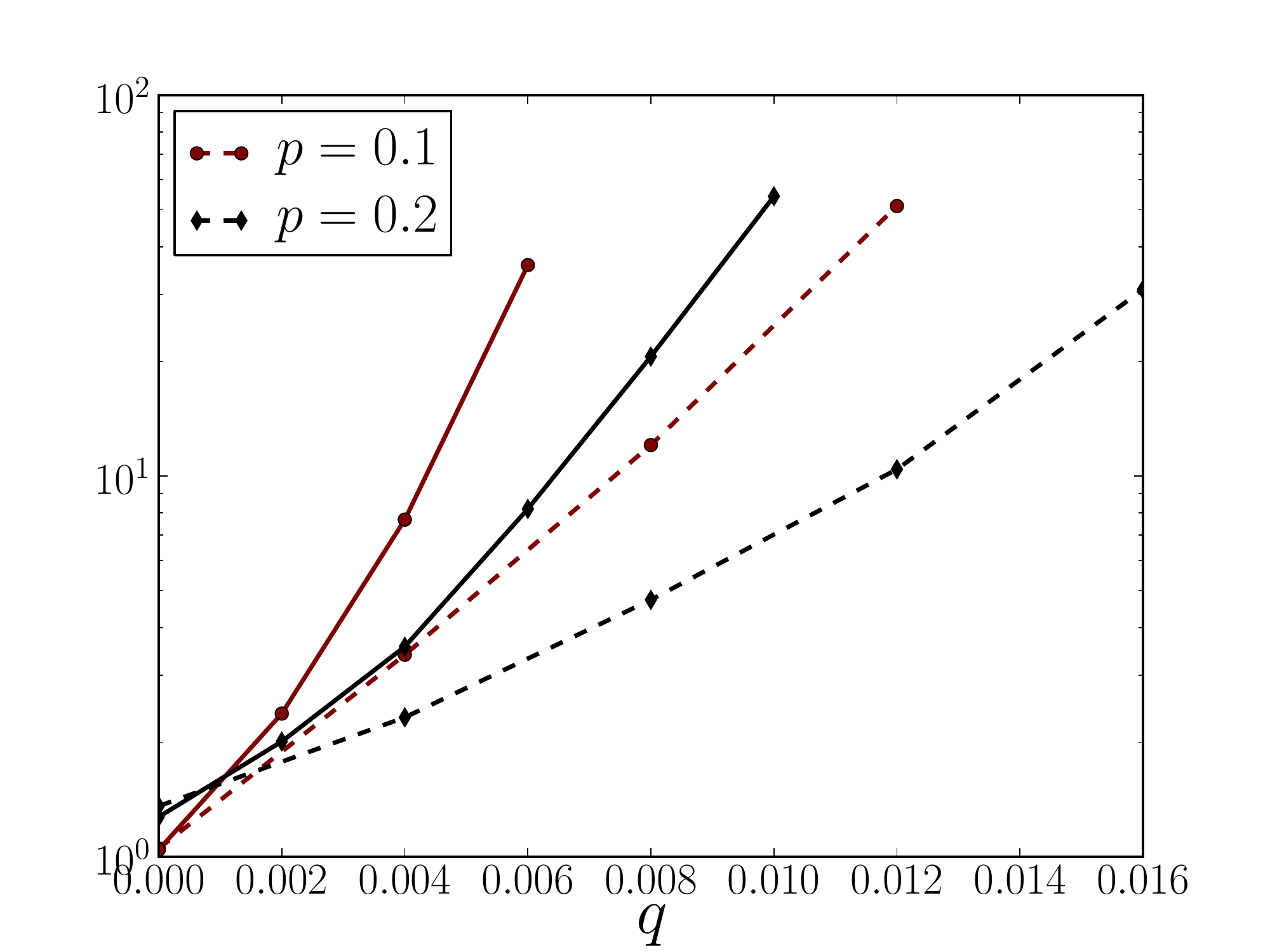}
\label{Smallp:a}
}
\hspace*{-0.2 in}
\subfigure[$S_1$]{
\includegraphics[width = 0.3\textwidth, keepaspectratio = true]{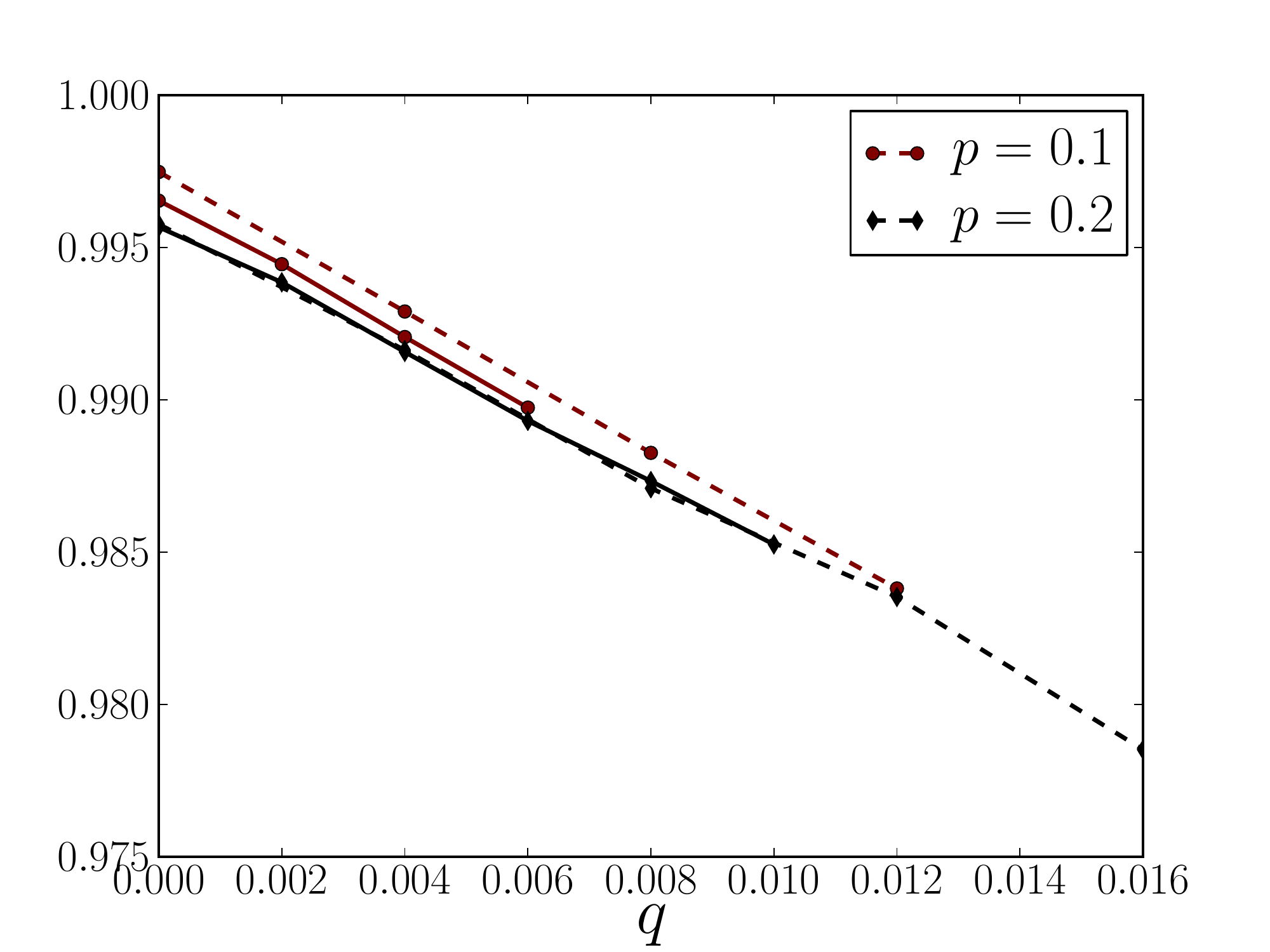}
\label{Smallp:b}
}
\hspace*{-0.2 in}
\subfigure[$N_c$]{
\includegraphics[width = 0.3\textwidth, keepaspectratio = true]{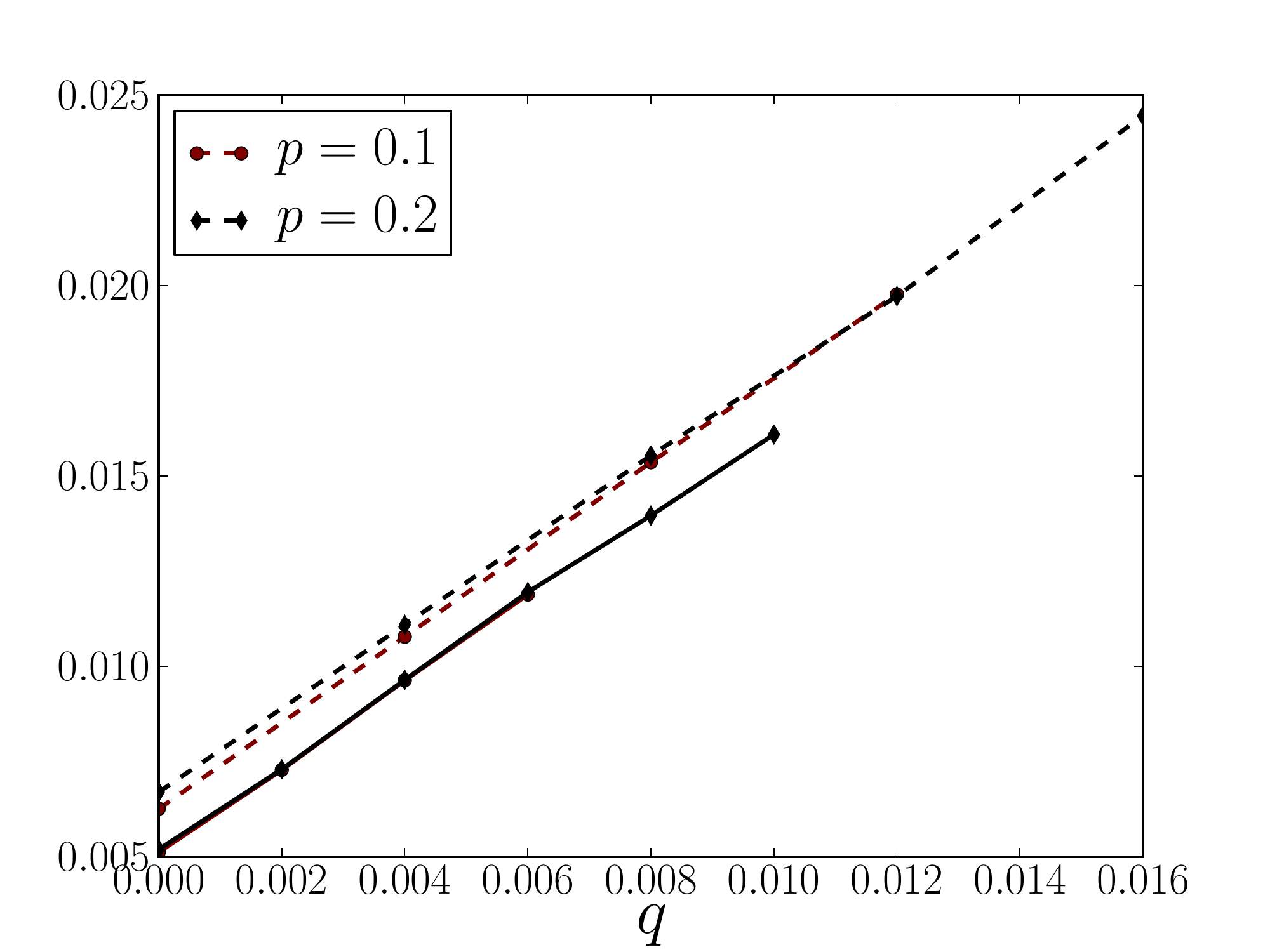}
\label{Smallp:c}
}
\caption[]{Characteristic time $\tau$, and the relative size of largest component ($S_1$) and relative number of components ($N_c$) as averaged over an ensemble in the topologically absorbing state, for small $p < p_c(q = 0)$. Solid lines $N = 500$, dashed lines $N = 250$.}
\label{Smallp}
\end{figure*}

\begin{figure*}
\hspace*{-0.2 in}
\subfigure[$\tau/N$]{
\includegraphics[width = 0.4\textwidth, keepaspectratio = true]{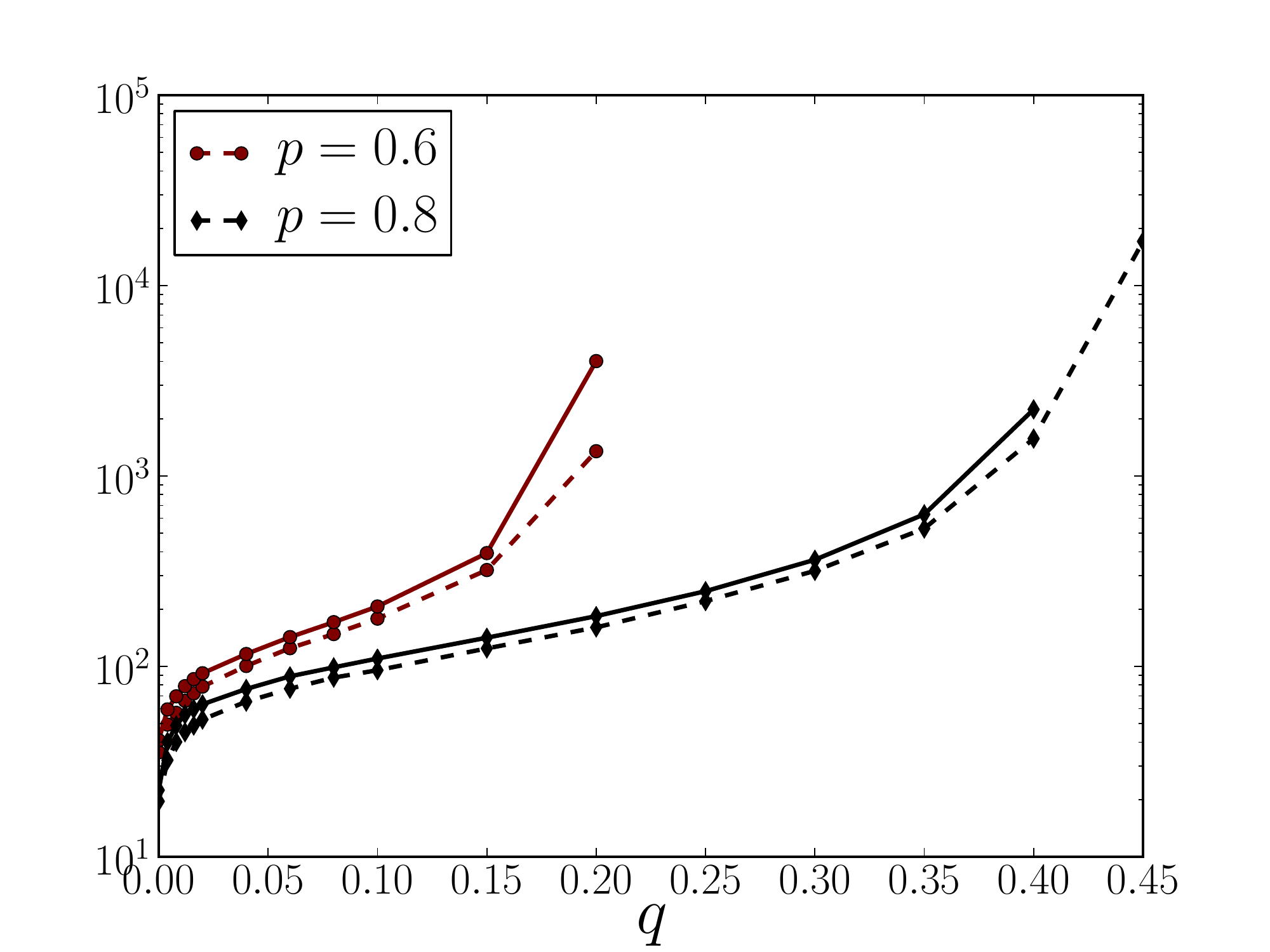}
\label{Largep:a}
}
\hspace*{-0.2 in}
\subfigure[Topologically absorbing state statistics]{
\includegraphics[width = 0.4\textwidth, keepaspectratio = true]{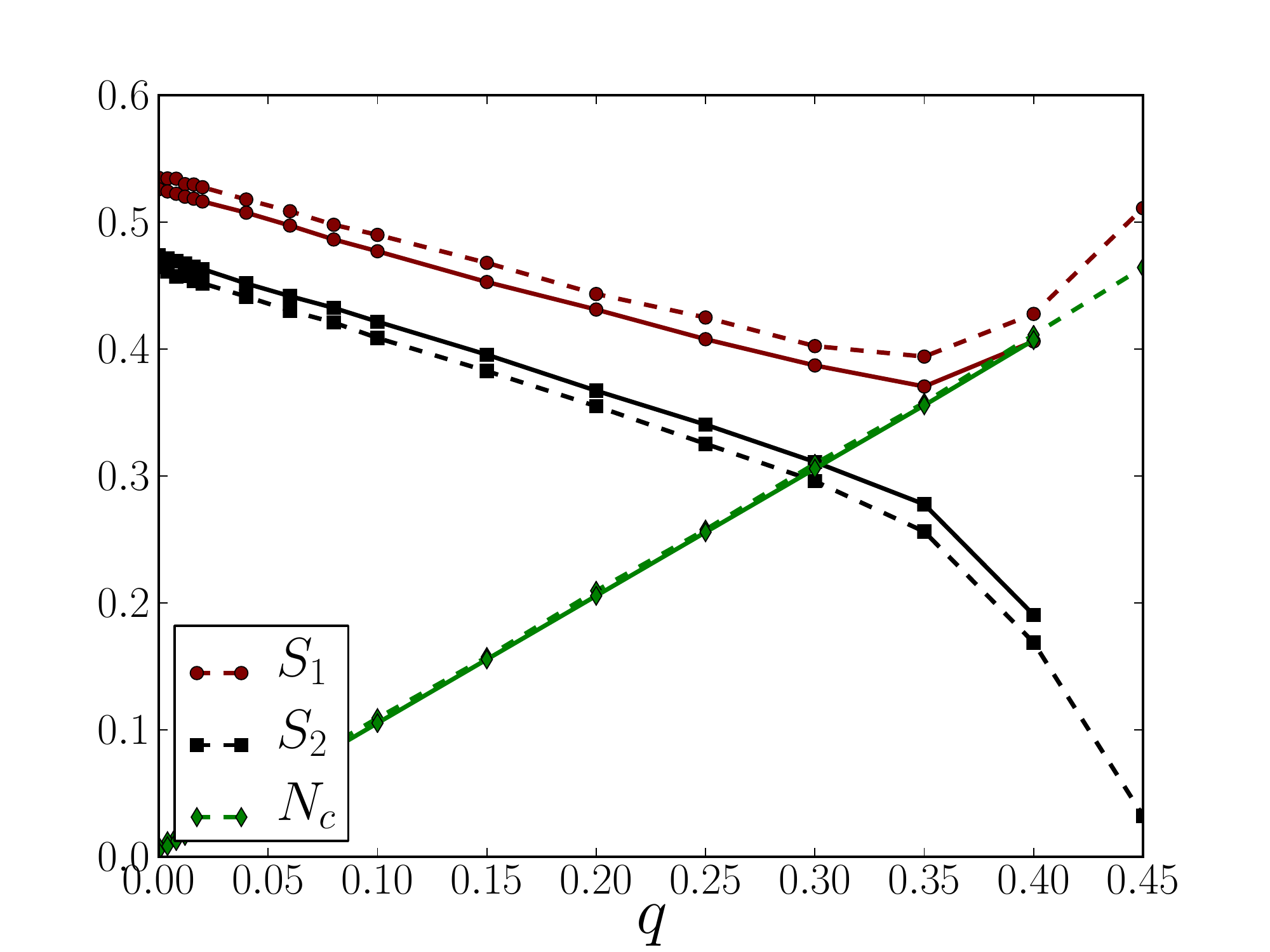}
\label{Largep:b}
}
\caption[]{Characteristic time $\tau$, and the averages in the topologically absorbing states, for large $p > p_c$.
Solid lines $N = 500$, dashed lines $N = 250$. Statistics were computed for progressively
increasing $q$ values, and the trends stop when the lifetime of realisations began to exceed an arbitrarily large computational 
limit ($10^5*N$).~\ref{Largep:b}: Average relative size of largest ($S_1$) and second largest ($S_2$) component, and 
relative number of components ($N_c$) at $p = 0.8$.}
\label{Largep}
\end{figure*}

Figure~\ref{Smallp:a} shows $\tau/N$ for small $p$, while the fraction $q$ of nodes targeted with 
noise increases. For any $q > 0$ shown, $\tau/N_1 > \tau/N_2$ if $N_1 > N_2$, which means the increase
of $\tau$ is superlinear with $N$. The same is observed when crossing into region A
by increasing $q$ at \emph{higher} values of $p$ (fig.~\ref{Largep:a}): after 
some critical $q$ the ratio of $\tau/N$ taken at two different $N$ begins to increase
significantly. The implication then
for region A is that infinite systems will take infinite time to reach the absorbing state, 
and hence that region A is an active region. 
We now consider how the being in region A affects the statistics of the topologically
absorbing configurations. When entering region A at small $p$, the relative size of the largest
component $S_1$ goes down linearly with $q$ (fig.~\ref{Smallp:b}), while the number of connected
components $N_c$ relative to system size $N$ increases with $q$ (fig.~\ref{Smallp:c}). The size of the second largest component
is negligible (given the largest component takes up almost all the nodes), and is not shown. 
This demonstrates that for small $p$, at least for the small values of $q$
that were considered, systems in region A freeze in one giant component and around 
$qN$ isolated nodes. Figure~\ref{Largep:b} shows the corresponding statistics
when approaching region A from region B, at large $p$ values. In region B
networks freeze into two large components and $qN$ isolated nodes. As 
$q$ increases the sizes of these components decrease accordingly, until
$S_1$ begins to increase, and $S_2$, the size of the second largest component,
decreases even faster. 
For example, when $p = 0.8$, $q = 0.45$, a network
of $N = 250$ nodes will, on average, freeze with the largest component
having half the nodes (all in the same state), the second largest component
containing around 8 nodes (probably in a different state), and the rest
of the network being isolated nodes and occasional node pairs as a finite-size effect.
The qualitative changes in the topological 
statistics and the characteristic times occur at the same $q$ range.
An $S_2$ of around zero is associated with the A region, and so we
use the difference $\Delta S = S_1 - S_2$ to denote the appearance of the 
A region: $\Delta S$ will stay constant until it begins to grow until 
it reaches a value around $S_1$. The finite-size fragmentation transition
can therefore be defined in terms of smallest/largest parameter values
when $\Delta S$ traverses some boundary. 

\section{Pair-approximation for Targeted Noise} 
\label{sec:pair_approximation_for_targeted_noise}
We present a step-by-step derivation of the difference equations governing the densities of different link kinds under Targeted Noise.\\
Our model consists of a random regular network of $N$ nodes, $L$ links, and average node degree $\mu$. Each node has a type $n_t$ that for convenience we call normal, $n$, and `noisy', or random, $r$ (thus $n_t \in \{n,r\}$). This type is assigned to the node at the start and does not change. We will use the variables $x$ and $y$ when referring to node types, so $x$ and $y$ take values $n$ or $r$. This partitions the links into ones that join together two $n$ nodes, two $r$ nodes, and an $n$ and an $r$ node. These three link \emph{types} can be written $xy \in \{nn, nr, rr\}$ ($rn$ is identified with $nr$). Further, since nodes can be in one of two states, we call a link active if it joins nodes currently in different states, and frozen otherwise. There are therefore six \emph{kinds} of links, depending on both the type and states of the nodes at both ends. Let the state vector that gives the density, i.e. the number of such links normalised by $L$, of each link \emph{kind} be $\boldsymbol{\rho}(t) = (\rho_n^a(t), \rho_n^f(t), \rho_r^a(t), \rho_r^f(t), \rho_m^a(t),\rho_ m^f(t))$, where $\rho_n^{a/f}$ is the density of active/frozen $nn$ links, $\rho_r^{a/f}$ the density of $rr$ links, and $\rho_m^{a/f}$ that of `mixed' $nr$ links. The asymptotic behaviour of $\boldsymbol{\rho}(t)$ therefore gives the asymptotic and topologically absorbing states.\\

\begin{figure*}
\hspace*{-0.2 in}
\includegraphics[width = 0.9\textwidth]{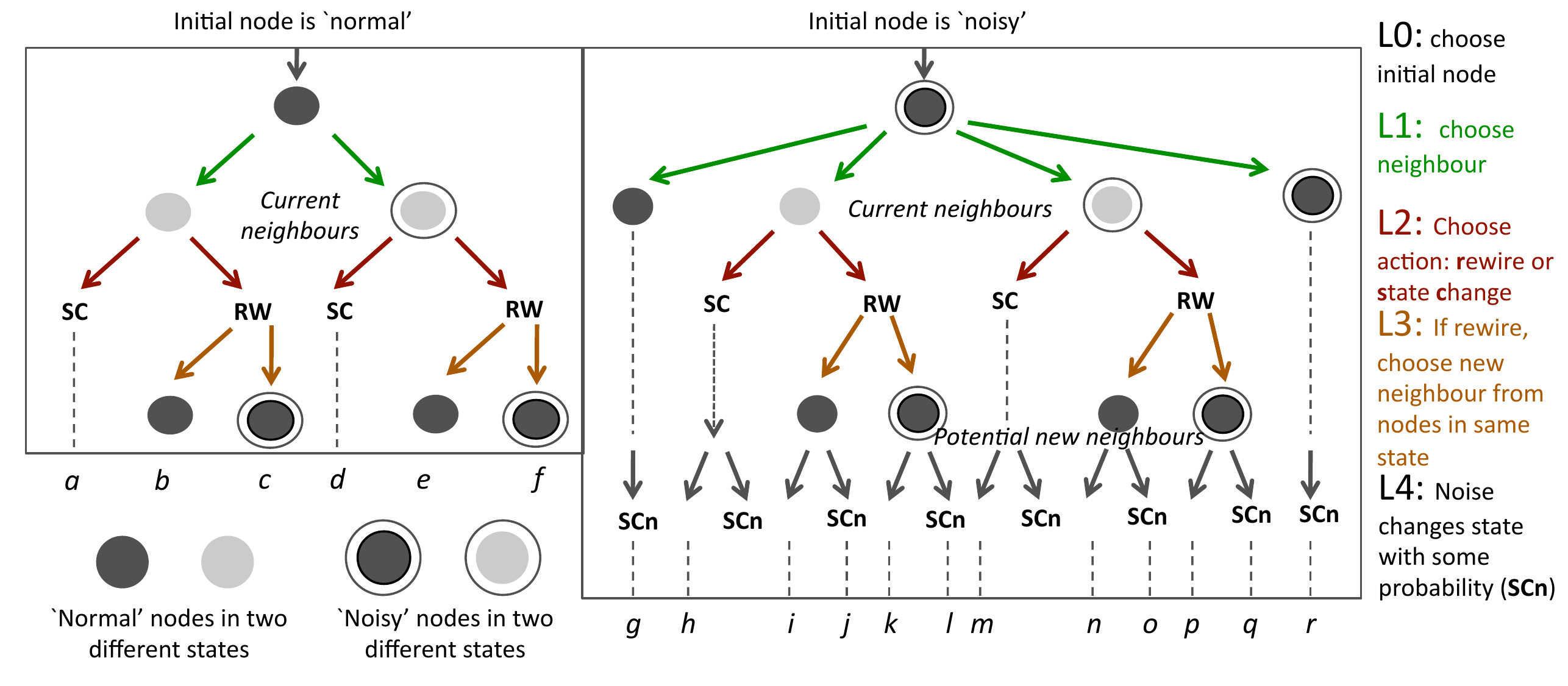}
\caption[]{Targeted Noise update scheme, showing only updates (marked $a-r$) that would effect some system change (topological or state-wise).}
\label{updates}
\end{figure*}

We now consider the possible updates and their effect on the numbers of the different link kinds. The schema in fig.~\ref{updates} shows the 
different ways a single update can proceed, ignoring those that would have no effect on the relative number of the different link kinds. The change within each time step to each link kind can be approximated by gathering the respective contributions from the 18 different ways an update can proceed, and weighing them with the probability of seeing the particular updates. The probabilities for the update stages L0 to L4, as well as the resulting contributions to the (unnormalized) differences in the elements of $\boldsymbol{\rho}$, are shown in table \ref{Table1}. Note that we first express the difference in absolute numbers, and then change the expression to work with the densities of links. 

	\begin{table}[h!]
  \begin{tabular}{|l||c|c|c|c|c | c|}
    \hline
&\multicolumn{6}{c|}{\textbf{Updates with `normal' initial node}} \\[5pt] \hline
	  & a & b & c & d & e & f \\ \hline
	Probabilities \\ \hline
	\, \textbf{L0} & $(1-q)$ & $(1-q)$ & $(1-q)$ & $(1-q)$ & $(1-q)$ & $(1-q)$ \\ 
	\, \textbf{L1} & $P(nn^a)$ & $P(nn^a)$ & $P(nn^a)$ & $P(nr^a)$ & $P(nr^a)$ & $P(nr^a)$ \\ 
	\, \textbf{L2} & $(1-p)$ & $p$ & $p$ & $(1-p)$ & $p$ & $p$ \\ 
	\, \textbf{L3} & n/a & $(1-q)$ & $q$ & n/a & $(1-q)$ & $q$ \\ 
    \hline 
	Contributions \\ \hline
	\, $nn^a$ & $-Q(nn^a)$ & $-1$ & $-1$ & $-Q(nn^a)$ 	&	   &\\ 
	\, $nn^f$ & $Q(nn^a)$ & $+1$  & 	 & $Q(nn^a)$ 	& $+1$ &\\ 
	\, $nr^a$ & $-Q(nr^a)$ &       & 	 & $-Q(nr^a)$ 	& $-1$ & $-1$\\ 
	\, $nr^f$ & $Q(nr^a)$ &        & $+1$ & $Q(nr^a)$  	&	   & $+1$\\ 
	\hline
  \end{tabular}

\hspace*{1in}

  \begin{tabular}{|c||c|c|c|c|c|c|c|c|c|c|c|c|}
    \hline
&\multicolumn{12}{c|}{
\textbf{Updates with `noisy' initial node}} \\[5pt] \hline
	  & g & h & i & j & k & l & m & n & o & p & q & r \\ \hline
	\hline
	\textbf{L0} & $q$ & $q$ & $q$ & $q$ & $q$ & $q$ & $q$ & $q$ & $q$ & $q$ & $q$ & $q$ \\ 	
	\textbf{L1} & $P(rn^f)$ & $P(rn^a)$ & $P(rn^a)$ & $P(rn^a)$ &  $P(rn^a)$ & $P(rn^a)$ &
		 $P(rr^a)$ & $P(rr^a)$ & $P(rr^a)$ & $P(rr^a)$ &  $P(rr^a)$ & $P(rr^f)$ \\
	\textbf{L2} & &$(1-p)$ & $p$ & $p$ &  $p$ & $p$ & $(1-p)$ & $p$ & $p$ & $p$ & $p$ &  \\ 
	\textbf{L3} & & & $(1-q)$ & $(1-q)$ & $q$ & $q$ &
	& $(1-q)$ & $(1-q)$ & $q$ & $q$ & \\ 
	\textbf{L4} & $\epsilon/2$ & $(1-\epsilon/2)$ &  $(1-\epsilon/2)$ & $\epsilon/2$ &  $(1-\epsilon/2)$ & $\epsilon/2$ &
	$(1-\epsilon/2)$ & $(1-\epsilon/2)$ & $\epsilon/2$ & $(1-\epsilon/2)$ & $\epsilon/2$ & $\epsilon/2$ \\ 
    \hline \hline
	$rr^a$ & $-Q(rr^a)$ & $-Q(rr^a)$ & & $-Q(rr^a)$& & $1-Q(rr^a)$ &
	$-Q(rr^a)$ & $-1$ & $-Q(rr^a)$ & $-1$ & $1-Q(rr^a)$ & $-Q(rr^a)$ \\ 
	$rr^f$ & $Q(rr^a)$ & $Q(rr^a)$ & & $Q(rr^a)$ & $+1$ & $Q(rr^a)$ &
	$Q(rr^a)$ & & $Q(rr^a)-1$ & $+1$ & $Q(rr^a)-1$ & $Q(rr^a)$\\	
	$nr^a$ & $-Q(rn^a)$ & $-Q(rn^a)$ & $-1$ & $1-Q(rn^a)$ &	$-1$ & $-Q(rn^a)$ &
	$-Q(rn^a)$ &  & $1-Q(rn^a)$ & & $-Q(rn^a)$ & $-Q(rn^a)$ \\ 
	$nr^f$ & $Q(rn^a)$ &	$Q(rn^a)$ & $+1$ & $Q(rn^a)-1$ & & $Q(rn^a) - 1$ &
	$Q(rn^a)$ & $+1$ &  $Q(rn^a)$ & & $Q(rn^a)$ & $Q(rn^a)$ \\

	\hline
  \end{tabular}
\caption{Possible updates of the Targeted Noise Coevolving Voter Model, for arbitrary noise intensity $\epsilon$, targeted fraction $q$,
and rewiring probability $p$. Contribution given in absolute numbers of links.}
\label{Table1}
\end{table}

To give the expressions for the terms appearing in table \ref{Table1} we need to consolidate notation and make some (further) approximations.
Since the number of links $L$ is conserved, there are $L\boldsymbol{\rho_i}(t)$, $i \in \{1,2..,6\}$ links of the $i^{\text{th}}$ kind at any time $t$. We define the probabilities L0-L4 and the contributions as follows:\\

\textbf{Probabilities}
\begin{itemize}
	\item{\textbf{L0}}: The probability of choosing an initial node of a certain type.\\
	Since initial nodes are chosen randomly, the probability of choosing a `normal' node
	initially is just the fraction of such nodes, which is $(1-q)$. The probability of choosing
	a `noisy' initial node is therefore $q$.
	\item{\textbf{L1}}: The probability of choosing a certain link type that results
	from randomly picking a neighbour given the initial node had already been chosen.\\
	 Clearly if for example a `normal' node was chosen initially, then the only links
	that can be picked are $nn$ and $nr$, and not the $rr$-type links.
	In the mean-field spirit we take the probability to pick a particular link kind 
	to be given by the relative number of such links among all the links
	of the node, which we take as representative of that node type.
	Let $P(nn^a)$ the probability to pick a $nn^a$ link given initial node is `normal', 
	$P(rn^a)$ is that of picking a $nr^a$ link given initial node is `random', etc.
	Consider $P(nn^a)$. The number of $nn^a$ links a `normal' node has is $\approx$ $\frac{2 L\rho_n^a}{(1-q)N}$,
	and the total number of links it has is $\approx \frac{2L\rho_n^a+ 2L \rho_n^f}{(1-q)N} + \frac{L\rho_m^a + L \rho_m^f}{(1-q)N}$,
	and hence
	\begin{equation}
		P(nn^a) \approx \frac{2\rho_n^a}{2(\rho_n^a+\rho_n^f)+\rho_m^a+\rho_m^f}.
	\end{equation}
	Similarly,
	\begin{align}
		\begin{split}
		& P(rr^a) \approx \frac{2\rho_r^a}{2(\rho_r^a+\rho_r^f)+\rho_m^a+\rho_m^f} \\
		& P(nr^a) \approx \frac{\rho_m^a}{2(\rho_n^a+\rho_n^f)+\rho_m^a+\rho_m^f} \\
		& P(rn^a) \approx \frac{\rho_m^a}{2(\rho_r^a+\rho_r^f)+\rho_m^a+\rho_m^f} \\
		\end{split}
	\end{align}
	\item{\textbf{L2}}: Probability to rewire or change state.\\ 
	Given by the model, and only applicable if an active link was chosen.
	\item{\textbf{L3}}: Probability to rewire to a `normal' or `noisy' node.\\
	Only applicable when performing rewiring. We assume the TD limit applies, so that
	there will \emph{always} be free nodes in the same state as the initial node,
	to rewire to. We approximate the densities of `normal' nodes as $(1-q)$, and
	`noisy' as $q$, so disregarding the small correction the selection of an initial 
	node would make. 
	\item{\textbf{L4}}: Probability to change state due to external noise, $\epsilon/2$.\\
	The node stays the same state with probability $(1 - \epsilon/2)$. Only applicable if the initial node is `noisy'. 
\end{itemize}

\textbf{Contributions}
Suppose that in a system of $N$ nodes, of which a fraction $q$ are `noisy', a random
node changes its state. This will lead active links connected to this node to become frozen, 
and vice versa. Depending on the type of node, we can approximate the contributions such 
a change would make to the difference in the absolute number of  each of the 
active/frozen ($a/f$) links kinds ($nn$, $rr$ and $nr$).\\
If the node is `normal', then changing its state will only affect the $nn$ and $nr$ links (
i.e. links joining together `normal' nodes, and links between `normal' and `noisy' nodes).
Just as before, let $L\rho_n^{a/f}$ and $L\rho_m^{a/f}$ be the number of, respectively, $nn$ and $nr$
links, either active or frozen. Then using the mean-field approach, each `normal' node 
will have $2L\rho_n^{a/f}/(1-q)N$ active/frozen $nn$ links, and $L\rho_m^{a/f}/(1-q)N$ active/frozen
$nr$ links. After the state change, the active links become frozen, and vice versa. Hence
for example the difference to the number of active $nn$ links would be $2L(\rho_n^{f}-\rho_n^{a})/(1-q)N$.
The four base formulas are: 
\begin{align*}
	\begin{split}
	& Q(nn^a) = \frac{2L(\rho_n^a - \rho_n^f)}{(1-q)N} \\
	& Q(nr^a) = \frac{L(\rho_m^a - \rho_m^f)}{(1-q)N} \\
	& Q(rn^a) = \frac{L(\rho_m^a - \rho_m^f)}{qN} \\
	& Q(rr^a) = \frac{2L(\rho_r^a - \rho_r^f)}{qN} \\
	\end{split}
\end{align*}

Since $L = N\mu/2$, let $N_n = 2(1-q)/\mu$ and $N_r = 2q/\mu$. Then
\begin{align*}
	\begin{split}
	& Q(nn^a) = \frac{2(\rho_n^a - \rho_n^f)}{N_n} \\
	& Q(nr^a) = \frac{(\rho_m^a - \rho_m^f)}{N_n} \\
	& Q(rn^a) = \frac{(\rho_m^a - \rho_m^f)}{N_r} \\
	& Q(rr^a) = \frac{2(\rho_r^a - \rho_r^f)}{N_r} \\
	\end{split}
\end{align*}
$Q(nn^a)$ is the change in the number of frozen $nn$ link as result of a `normal' node changing 
state, $Q(nr^a)$ is the corresponding change in the number of frozen $nr$ links. Conversely,
if a `noisy' node changes state, $Q(rr^a)$ would be the resulting difference in the number of 
frozen $rr$ links, and $Q(rn^a)$ that in the frozen $nr$ links. Note that $Q(rn^a)$ is not
in general equal to $Q(nr^a)$.\\
	
	The equations for the difference in the six link densities are obtained from
	summing the respective contributions from table \ref{Table1}, and normalising
	by $1/L$ and $N$ (to obtain the difference in a unit of time, defined as having $N$ updates).
	For $\epsilon = 1$ this gives the system of equations found in the main text of the paper:
	
	\begin{align*} \label{allequations}
		\begin{split}
		& \Delta \rho_n^a = 2/\mu \left[  (1-p)(q-1)P(nr^a)Q(nn^a)+(1-p)(q-1)P(nn^a)Q(nn^a)+(q-1)p P(nn^a) \right]\\
		& \Delta \rho_n^f = 2/\mu \left[-(q-1)(P(nr^a)+P(nn^a))((1-q)p+(1-p)Q(nn^a))\right]\\
		& \Delta \rho_r^a = 2/\mu \left[-\frac{q}{2}Q(rr^a)+\frac{q}{2}p(q-1)P(rr^a)+\frac{q^2}{2}p P(rn^a)\right]\\
		& \Delta \rho_r^f = 2/\mu \left[\frac{q}{2}Q(rr^a)+\frac{q}{2}p(q-1)P(rr^a)+\frac{q^2}{2}p P(rn^a)\right]\\
		& \Delta \rho_m^a = 2/\mu \left[-\frac{q}{2}Q(rn^a)-\frac{q^2}{2}p P(rn^a)+\frac{q}{2}p(1-q)P(rr^a)-(1-q)(p+(1-p)Q(nr^a))P(nr^a)-(1-q)(1-p)P(nn^a)Q(nr^a)\right]\\
		& \Delta \rho_m^f = 2/\mu \left[\frac{q}{2}Q(rn^a)-\frac{q^2}{2}p P(rn^a)+\frac{q}{2}p(1-q)P(rr^a) + (1-q)(pq+(1-p)Q(nr^a))(P(nr^a)+P(nn^a))\right].
		\end{split}
	\end{align*}


\bibliography{Bibliography}

\end{document}